\documentclass[lettersize,journal]{IEEEtran}
\usepackage{amsmath,amssymb,amsfonts}
\usepackage{algorithmic}
\usepackage{algorithm}
\usepackage{array}
\usepackage{textcomp}
\usepackage{stfloats}
\usepackage{url}
\usepackage{verbatim}
\usepackage{graphicx}
\usepackage{cite}
\usepackage{wrapfig}

\usepackage{hyperref}
\usepackage{gensymb}
\usepackage{tabularx}
\usepackage{longtable}
\usepackage{float}
\usepackage{soul}
\usepackage{multirow}

\usepackage{enumitem}
\usepackage{booktabs}
\usepackage{hhline}
\usepackage{boldline} 
\usepackage{ragged2e}
\usepackage{mathtools}
\usepackage{breqn}
\usepackage{color}

\newcolumntype{P}[1]{>{\centering\arraybackslash}p{#1}}
\newcolumntype{M}[1]{>{\arraybackslash}m{#1}}
\usepackage{subcaption}

\begin{document}

\title{Noncontact Respiratory Anomaly Detection Using Infrared Light-Wave Sensing}

\author{Md Zobaer Islam, Brenden Martin, Carly Gotcher, Tyler Martinez, John F. O'Hara, \IEEEmembership{Senior Member, IEEE}, \\and Sabit Ekin, \IEEEmembership{Senior Member, IEEE}
\thanks{2168-2291 © 2024 IEEE. Personal use is permitted, but republication or redistribution requires IEEE permission. See https://www.ieee.org/publications/rights/index.html for more information.}
\thanks{
This work was supported by the National Science Foundation under Grants 2008556, 2323301 and 2336852. (\textit{Corresponding author: Md Zobaer Islam, Sabit Ekin.})}
\thanks{Md Zobaer Islam is with the Department of Radiology, University of North Carolina at Chapel Hill, Chapel Hill, North Carolina, USA, and the School of Electrical \& Computer Engineering, Oklahoma State University, Oklahoma, USA (e-mail: zobaer\_islam@med.unc.edu, zobaer.islam@okstate.edu)
}
\thanks{Brenden Martin, Carly Gotcher, Tyler Martinez, and John F. O'Hara are with the School of Electrical \& Computer Engineering, Oklahoma State University, Oklahoma, USA (e-mail: brenden.martin, carly.gotcher, tyler.martinez, oharaj \{@okstate.edu\})
}
\thanks{Sabit Ekin is with the Departments of Engineering Technology, and Electrical \& Computer Engineering, Texas A\&M University, College Station, Texas, USA (e-mail: sabitekin@tamu.edu)
}
\thanks{Digital Object Identifier 10.1109/THMS.2024.3381574}
}

\markboth{ARXIV VERSION UPLOADED: PUBLISHED IN IEEE TRANSACTIONS ON HUMAN-MACHINE SYSTEMS}%
{Shell \MakeLowercase{\textit{et al.}}: A Sample Article Using IEEEtran.cls for IEEE Journals}


\maketitle

\begin{abstract}
Human respiratory rate and its pattern convey essential information about the physical and psychological states of the subject. Abnormal breathing can indicate fatal health issues leading to further diagnosis and treatment. Wireless light-wave sensing (LWS) using incoherent infrared light shows promise in safe, discreet, efficient, and non-invasive human breathing monitoring without raising privacy concerns. The respiration monitoring system needs to be trained on different types of breathing patterns to identify breathing anomalies. 
The system must also validate the collected data as a breathing waveform, discarding any faulty data caused by external interruption, user movement, or system malfunction. To address these needs, this study simulated normal and different types of abnormal respiration using a robot that mimics human breathing patterns. Then, time-series respiration data were collected using infrared light-wave sensing technology. Three machine learning algorithms, decision tree, random forest and XGBoost, were applied to detect breathing anomalies and faulty data. Model performances were evaluated through cross-validation, assessing classification accuracy, precision and recall scores. The random forest model achieved the highest classification accuracy of 96.75\% with data collected at a 0.5\,m distance. In general, ensemble models like random forest and XGBoost performed better than a single model in classifying the data collected at multiple distances from the light-wave sensing setup.
\end{abstract}

\maketitle

\section{Introduction}
\label{sec:intro}
\IEEEPARstart{T}{he} respiratory patterns of human beings can serve as a reflection of both their physical and psychological states. By monitoring these patterns, it is possible to increase general health awareness and detect anomalous breathing before more serious health complications arise. Anomalous breathing may result from respiratory illnesses such as asthma, obstructive sleep apnea, chronic bronchitis, chronic obstructive pulmonary disease, emphysema, and COVID-19. Also, it can be a symptom of unstable mental conditions including stress, panic, anxiety, fatigue, and anger. Although respiratory rate and pattern are considered valuable indicators of worsening health conditions, medical literature indicates that the assessment of respiratory rate and quality are frequently erroneous or completely omitted in clinical settings~\cite{elliott2016respiratory, hogan2006don}. The manual counting of breathing rate and subjective evaluation of its quality are prone to errors and not suitable for continuous monitoring~\cite{van2008missed}. Automated breathing monitoring solutions generally employ contact-based or wearable sensors~\cite{grover2008automated}. When a person is aware that their breathing is being monitored, they may undergo a shift in breathing pattern from their typical signature. Furthermore, contact-based approaches may be unsuitable for patients who are either too young, such as infants in a neonatal intensive care unit (NICU), or too ill, such as burn unit patients~\cite{Smith2016}.  Additionally, contact-based respiration monitoring may spread contagious diseases such as COVID-19. Hence, there is a pressing need for noncontact, discreet respiration monitoring techniques.

Noncontact vitals monitoring using radio frequencies (RF) has been the subject of research and development for several years. Various studies have utilized received signal strength and channel state information from WiFi networks to detect, estimate, and monitor human breathing~\cite{Abdelnasser2015,bao2022wi,guo2023breatheband}. Similarly, remote breathing monitoring has been investigated using different types of radars~\cite{saeed2021wireless,SanKocur2017,Purnomo2021}. However, RF-based respiration monitoring approaches are susceptible to electromagnetic interference with signals from nearby devices~\cite{LAPINSKY2006267, MARIAPPAN2016727}. Also, continuous exposure to RF signals can be detrimental to the human body~\cite{TAKI200140, gye2012effect}. Furthermore, the WiFi-based approaches usually require a receiver device in contact with the chest, thus not qualifying as fully noncontact methods~\cite{Abdelnasser2015}. In addition to RF signal-based methods, researchers have explored alternative approaches utilizing optical imaging-based techniques to monitor respiration and other physiological signals~\cite{zhang2023recent}. These methods primarily involve the use of recorded videos or images through RGB~\cite{ Brieva2020,romano2021, Hwang2021,tan2023lightweight}, depth sensing~\cite{s21041135, kempfle123,wang2020unobtrusive}, and thermal infrared~\cite{Jagadev2020, chen2017hht, shu2022} cameras for extracting information related to breathing. However, these approaches can potentially raise privacy concerns among users because the subjects' image data is discreetly captured. Moreover, post-processing of video and image data is computationally more expensive than processing one-dimensional time-series data~\cite{Abuella2020}.

Noncontact vitals monitoring using light-wave sensing may prove superior to existing technologies because of the safe, ubiquitous and harmless nature of incoherent light as well as the absence of privacy issues that are common with camera-based approaches. Noncontact vitals monitoring using visible light has already been performed as a proof of concept that showed $>$94\% accuracy in breathing rate measurement~\cite{Abuella2020}. But visible light can be troublesome to subjects in dark environments, particularly during sleep. Therefore, more discreet alternatives, such as infrared (IR) light, are preferred, as IR light is not visible to the naked human eye. This study introduces a novel system model for human respiration monitoring and anomaly detection using IR-based light-wave sensing technology. We collected human-like respiration data in a controlled environment with high precision and repeatability. To identify anomalous breathing and faulty data, we applied machine learning algorithms to the handcrafted features extracted from the collected data. Therefore, the main contributions of this project are the development of the IR-based light-wave sensing system for noncontact respiration monitoring and the use of machine learning to smartly detect anomalous breathing and discard faulty data.

The remainder of this manuscript is organized as follows. Section~\ref{sec:related_works} includes an overview of related works from the literature on breathing anomaly detection using various sensing technologies and machine learning. Section~\ref{sec:human_breath_pattern} describes various human breathing patterns from the literature to be used as breathing classes for anomaly detection. 
Section~\ref{sec:resp_sensing} presents the system model, relevant theory and lock-in detection process used in this study. The details of hardware components, data collection and initial data processing are depicted in Section~\ref{sec:data_collect_and_proc}. Next, Section~\ref{sec:feature_ex} describes the handcrafted features used and their extraction process. Data classification process using the chosen machine learning algorithms are included in Section~\ref{sec:data_classification} and the results along with their interpretations are discussed in Section~\ref{sec:model_eval}. Finally, Section~\ref{sec:conclusion} presents the conclusions drawn from the whole effort and forecasts future research directions.

\section{Related Works}
\label{sec:related_works}
Researchers have been applying machine learning and deep learning techniques on human respiration data collected through various technologies for anomaly detection. Most of these efforts made use of handcrafted features to perform breathing data classification for anomaly detection. Some of the common categories of features used in the literature were statistical features from the data (mean, standard deviation, skewness, kurtosis, root mean-square value, range etc.), signal-processing based features (Fourier co-efficients, autoregressive integrated moving average co-efficients, wavelet decomposition, mel-frequency cepstral coefficients, linear predictive coding etc.), and respiration related features (breathing rate, amplitude, inspiratory time, expiratory time etc.)~\cite{Rehman2021,Ucar2017, Pegoraro2021, Purnomo2021,Fekr2016, Jagadev2020}. In some research efforts, deep neural networks were trained to recognize subtle features from breathing data before classification, thus making manual feature extraction redundant~\cite{Hwang2021, Kim2019, saeed2021wireless, Brieva2020, shu2022}.

Some past classification efforts involved one-class classification or outlier detection, as in~\cite{Pegoraro2021} where the model was trained using human breathing data in resting condition to predict if the person was exercising in new examples. Binary classification between normal breathing and apnea were performed in~\cite{Ucar2017} to detect obstructive sleep apnea. Multiclass breathing classification efforts considered different types of breathing anomalies like tachypnea, bradypnea, hyperpnea, hypopnea etc. and sometimes more complicated anomalies like Cheyne-Stoke's, Biot's and Apneustic breathing as separate classes~\cite{Jagadev2020, Kim2019, Fekr2016}. Most of these breathing patterns are explained in Section~\ref{sec:human_breath_pattern}. Data for these efforts were usually obtained from human volunteers who are generally unable to breathe using precise frequency, amplitude and pattern. Occasionally, data from patients with breathing disorders were utilized, but this approach had its limitations as well. This is because even the patients may not consistently exhibit abnormal breathing patterns which increases the risk of mislabeling the training data. In the current study, more reliable data were generated by using a programmable robot with precise human-like breathing capability. Various machine learning techniques were employed in the literature to classify breathing data, including decision tree, random forest, support vector machine, XGBoost, $K$-nearest neighbors, feedforward neural network, and logistic regression, among others. The performance of these models was assessed using different evaluation metrics such as confusion matrices, $K$-fold cross-validation, accuracy, precision, sensitivity (recall), specificity, F1-score etc.~\cite{Fekr2016,Ucar2017,Jagadev2020, Purnomo2021, Hwang2021}.

\section{Human Breathing Patterns}
\label{sec:human_breath_pattern}
Various types of human breathing patterns (both normal and abnormal) have been identified in the available literature. The seven major types of human breathing which will be addressed in this study are described as follows:

\begin{enumerate}[leftmargin=*]
\item \textbf{Eupnea}: This refers to regular human breathing with a uniform depth, rate and pattern. The specific depths and rates considered normal vary based on factors such as age and activity level. In adults, the typical resting breathing rate falls within the range of 12-20 breaths per minute (BPM)~\cite{BarbosaPereira2017,Fekr2016,Purnomo2021,Yuan2013}. Breathing depth is determined by measuring the movement of the rib cage and expressing it as a percentage of the maximum rib cage movement. Experimental findings indicate that healthy adults aged 20-39 exhibit a breathing depth of approximately 44$\pm$14\% or 30-58\% of the maximum rib cage movement~\cite{Parreira2010}.

\item \textbf{Apnea}: Temporary cessation of breathing is known as apnea~\cite{BarbosaPereira2017,Ali2021}. It often occurs during sleep which is known as sleep apnea.

\item \textbf{Tachypnea}: This is an anomalous breathing condition characterized by a faster breathing rate than usual~\cite{BarbosaPereira2017,Rehman2021,Ali2021,Jagadev2020,Moraes2019}. A breathing rate exceeding 20\,BPM is typically considered as tachypnea~\cite{Fekr2016}. 

\item \textbf{Bradypnea}: Bradypnea is defined as slow breathing~\cite{BarbosaPereira2017,Ali2021,Jagadev2020,Purnomo2021}. So, any breathing rate less than 12\,BPM can be considered as bradypnea~\cite{Fekr2016}.

\item \textbf{Hyperpnea}: Hyperpnea is a breathing pattern with increased depth of breathing at a normal rate~\cite{Moraes2019,Purnomo2021}.

\item \textbf{Hypopnea}: In hypopnea, the breathing becomes shallow with at least 50\% decrease in the regular air flow volume for $\geq$10 seconds~\cite{Leung2012,Weinreich2009,Ucar2017}.

\item \textbf{Kussmaul's breathing}: It refers to a condition where tachypnea and hyperpnea coexist, resulting in rapid, deep, and labored breathing. In Kussmaul's breathing, both the depth and rate of breathing exceed those of eupnea or normal breathing~\cite{BarbosaPereira2017,Rehman2021,Moraes2019,Yuan2013}. 
\end{enumerate}

\begin{figure}[!htbp]
 \centering
 \includegraphics[width=0.44\textwidth]{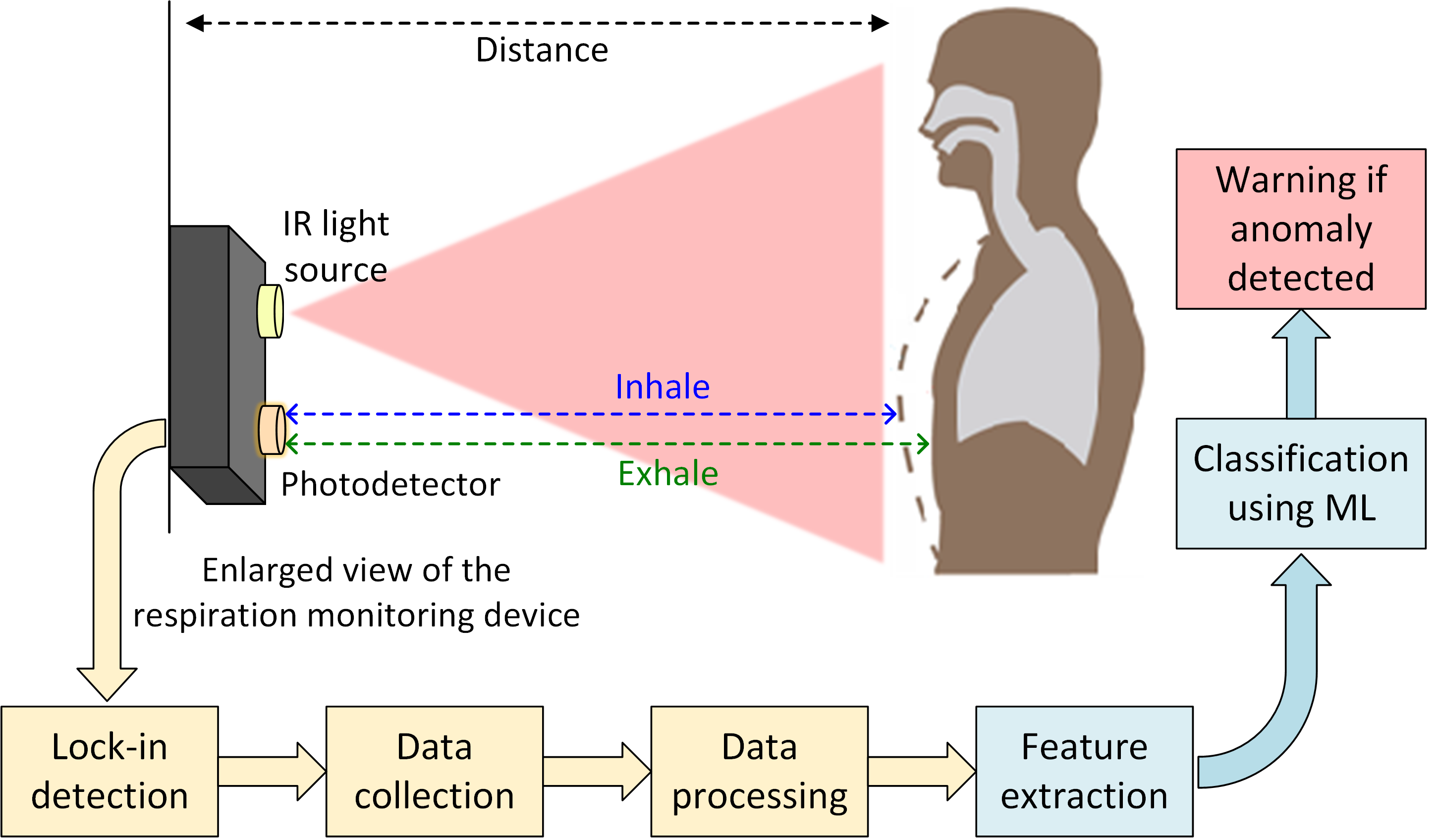}
 \caption{Proposed system model for respiratory anomaly detection using infrared (IR) light-wave sensing}
 \label{fig:sysmodel}
 \end{figure}

\section{Respiration Sensing}
\label{sec:resp_sensing}
\subsection{Theory of Operation}
The proposed system model for noncontact respiration monitoring using IR sensing and anomaly detection is illustrated in Fig.~\ref{fig:sysmodel}. It showcases the key components, including an IR light source and photodetector, a human subject positioned in front of them, and the functional blocks involved in the process. According to this model, modulated IR light is sent towards the chest of a human being. Light propagation from light emitting diodes (LED) is assumed to follow Lambertian propagation model~\cite{gfeller1979,pathak2015}. According to this model, if $P_t$ is the optical power transmitted from a point source, then power at distance $d$, is given by
\begin{align}
   \label{eq:Lambertian_Channel1}
    P_d=\frac{(n+1)A P_t}{2\pi d^\gamma}\cos^n(\phi)\cos(\theta) , \forall~\theta< \phi_{1/2},
\end{align}
where $A$ is the area intercepted, $\gamma$ is the empirical path-loss exponent, and $\phi$ and $\theta$ are irradiance and incident angles, respectively~\cite{Abuella2020}. $\phi_{1/2}$ is the half-power angle in the field of view of the light source. $n$ is the order of the Lambertian model and is given by $n=\frac{-\ln(2)}{\ln\{\cos(\phi_{\frac{1}{2}})\}}$.

Propagated optical power intercepts the subject's chest, part of which is absorbed by the clothing material while the rest is further modulated by the chest movement due to inhaling and exhaling and is scattered back to the photodetector. If $P_r$ is the scattered optical power received by the photodetector, $R_{pd}$ is the responsivity and $i_d$ is the dark current of the photodetector, then the generated photocurrent $i_{pd}=i_d+R_{pd}P_r$~\cite{detector2}. The transimpedance amplifier in the photodetector converts this current into voltage
\begin{align}
V_{sig}=g_{pd}\left(i_d+R_{pd}P_r\right),
\end{align}
where $g_{pd}$ is the transimpedance gain. This signal $V_{sig}$ is contaminated by noise from both the photodetector and the surrounding environment.

\subsection{Lock-in Detection}
When the signal of interest is buried in noise, lock-in detection technique can be used to detect the signal accurately. Hence, the output voltage from the photodetector ($V_{sig}$) is fed to the lock-in amplifier~\cite{lockin1} as input. In lock-in detection, the source signal (transmitted light) is modulated using a periodic signal of known frequency $\omega_r$ and amplitude $V_r$. The same signal is used as reference input of the lock-in  amplifier. The phase of this reference signal can be treated as zero. The lock-in amplifier generates an internal reference oscillator signal $V_{ref1} = V_r\sin (\omega_rt+\phi_r)$ that has a constant phase difference $\phi_r$ with the external reference signal. Assume that the signal to be measured ($V_{sig})$ is a sinusoidal wave with only one frequency $\omega_s$ and phase $\phi_s$ i.e. $V_{sig} = V_s\sin (\omega_st+\phi_s)$. The phase sensitive detector (PSD) of the lock-in amplifier mixes these two signals by multiplying them together to produce the output
\begin{equation}
\begin{split}
   V_{m1} = &\,\frac{V_sV_r}{2}[\cos\{\left(\omega_s-\omega_r\right)t+\left(\phi_s-\phi_r\right)\}-\\&\cos\{\left(\omega_s+\omega_r\right)t+\left(\phi_s+\phi_r\right)\}].
\end{split}
\end{equation}

$V_{m1}$ contains a sum and a difference frequency components. The frequencies $\omega_s$ and $\omega_r$ will be equal since the same signal was used as the reference and the source carrier. Thus, the difference frequency term reduces to a DC component proportional to the signal amplitude. A low-pass filter removes the sum frequency component from the signal. By denoting $\phi_s-\phi_r=\Delta\phi$, the filtered output becomes $V_x = \frac{V_sV_r}{2}\cos(\Delta\phi)$. To eliminate the phase dependency from the output voltage, a second internal reference signal $V_{ref2} = V_r\cos(\omega_rt+\phi_r)$ is used that is in quadrature to the first one.  A second PSD mixes $V_{sig}$ and $V_{ref2}$ and the mixed output is filtered similarly by low pass filter with the same specifications to yield the filtered output $V_y = \frac{V_sV_r}{2}\sin(\Delta\phi)$. $V_x$ and $V_y$ can be combined to find the magnitude $ R=\sqrt{V_x^2+V_y^2}=\frac{V_sV_r}{2}$ and phase $\theta=\tan^{-1}\left(\frac{V_y}{V_x}\right)=\Delta\phi$ after lock-in detection. Here, $R$ is proportional to the amplitude ($V_s$) of the signal to be measured. It can be scaled to the desired level by varying the sensitivity $S=\frac{V_{fs}}{G}$, where $V_{fs}$ is the full-scale voltage (genarally 10V) and $G$ is the overall gain of the lock-in amplifier, to produce
\begin{align}
R_{scaled} = \frac{V_{fs}V_r}{2S}V_s.
\end{align}

Since the signal is distorted by high-frequency noises, noise frequencies will similarly generate sum and difference frequencies in the mixed signal, but both components will be far from the DC component of interest and, hence, will be attenuated by the low pass filter. Filter bandwidth can be adjusted by varying the time constant, $\tau=\frac{1}{2\pi f_c}$ of the lock-in amplifier, where $f_c$ is the 3-dB cut-off frequency of the low pass filter. A narrowband filter will suppress most of the noises while keeping the signal to be detected intact as a true DC component.

\begin{figure}[!hbt] \centering
 
\begin{subfigure}[!hbt]{0.2\textwidth}
 \centering
\includegraphics[width=\textwidth]{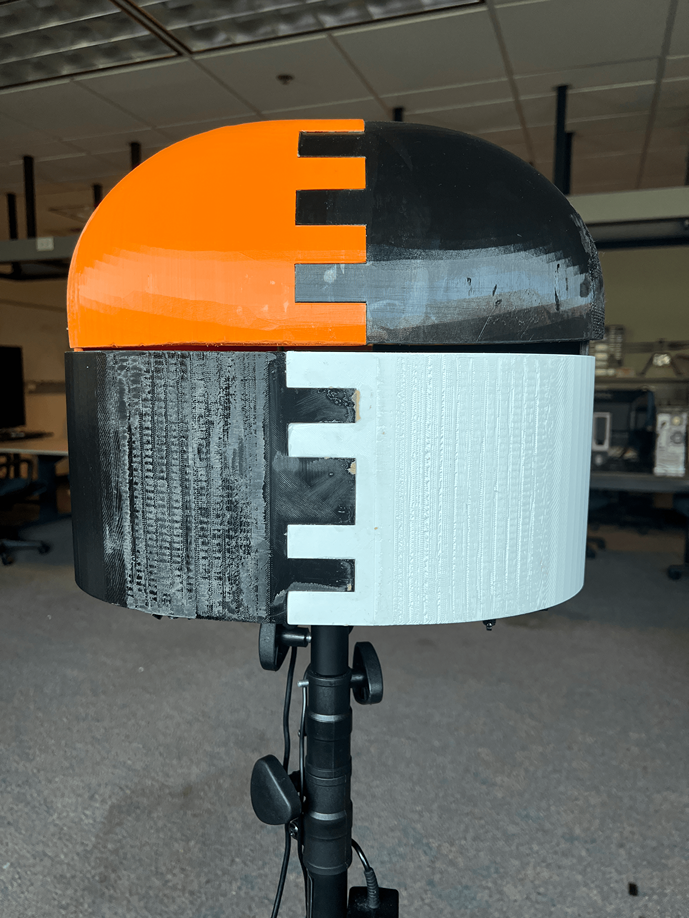}
\caption{External view.}
\label{fig:external}
\end{subfigure}
\hfill
\begin{subfigure}[!hbt]{0.2\textwidth}
 \centering
\includegraphics[width=\textwidth]{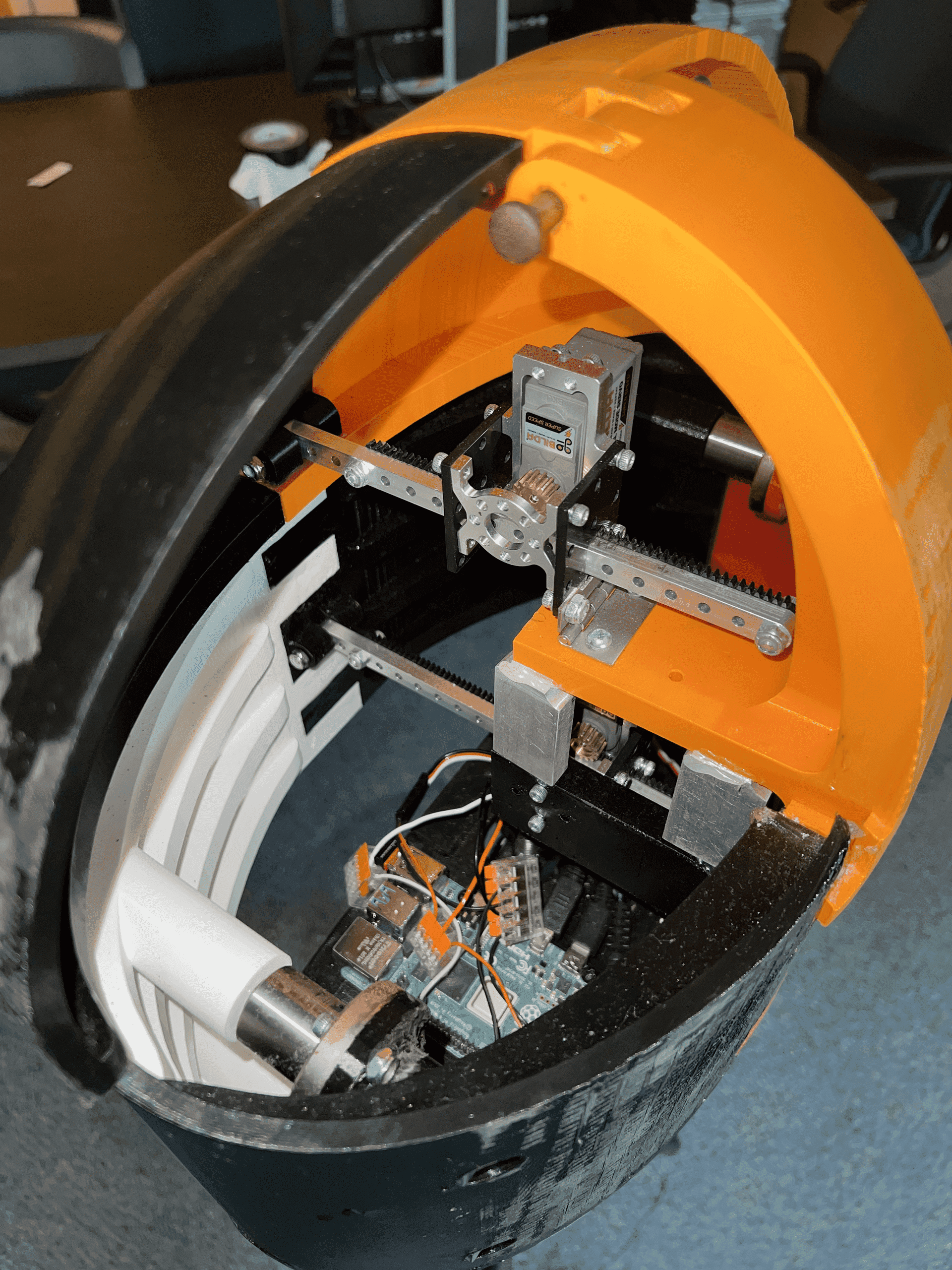}
\caption{Internal view.}
\label{fig:internal}
\end{subfigure}
\caption{The robot's chest-rib and motor assembly, operated by a Raspberry Pi.}
\label{fig:robot_views}
\end{figure}

\section{Data Collection and Processing}
\label{sec:data_collect_and_proc}

\subsection{Robot-simulated Respiration}
In order to collect respiration data in a controlled, precise and repeatable way, a robot was developed to emulate human respiration. In Fig.~\ref{fig:external}, the 3D-printed torso of the robot is presented, highlighting its physical structure. Fig.~\ref{fig:internal} showcases the internal hardware assembly of the robot, including servo motors, gears, racks, pinions, and other components. At the bottom of the assembly, a Raspberry Pi is positioned to control breathing rate, depth and pattern of the robot. Two Python programs were employed to control the robot. The first program enabled the robot to emulate multiple predefined breathing waveforms such as $\sin^2$, $\sin^4$ and $\sin^6$ patterns. The use of even powers of sine waves to incorporate harmonics beyond the fundamental breathing frequency has been argued for radar applications~\cite{Breathing_Model}. Among these patterns, $\sin^4$ and $\sin^6$ patterns resemble the waveform of human breathing more closely by representing the duration and intensity of inhalation and exhalation more accurately. The breathing rate could be adjusted from 0 to 50\,BPM with a maximum depth ranging from 0 to 30 mm (expressed as 0\% to 100\% in the program), both covering the full range of potential normal and abnormal breathing rates and depths in humans. We validated the length of chest displacement by measuring it with a ruler. The breathing rate was validated by manual counting and by identifying the frequency with the highest spectral amplitude in the Fourier transform of breathing data collected using Terabee time-of-flight sensor~\cite{terabee}. An initial chest position offset could also be set. A second control software could actuate arbitrary waveforms from a text file generated by MATLAB code.

For the present study on infrared sensing for detecting respiratory anomalies, respiration data with precise frequencies and depths were necessary to create labeled training data and evaluate performance on test data. The data included regular breathing and six different types of anomalous breathing. Since breathing is an involuntary activity, humans have limited control over their breathing rates and depths. Therefore, it is extremely challenging for humans to consistently breathe at various prescribed rates and depths for the purpose of collecting training data, unless they have received specialized training to do so. Using a breathing machine or robot eliminated the need to establish ground truth as the robot operator controlled the breathing parameters for each scenario. Moreover, the machine could consistently generate data for extended durations, resulting in a more comprehensive and reliable dataset. Therefore, the use of a robot allowed us to conduct research by finely adjusting various breathing parameters, as well as assessing the system's performance and limitations.

\begin{figure}[t]
 \centering
 \includegraphics[width=0.42\textwidth]{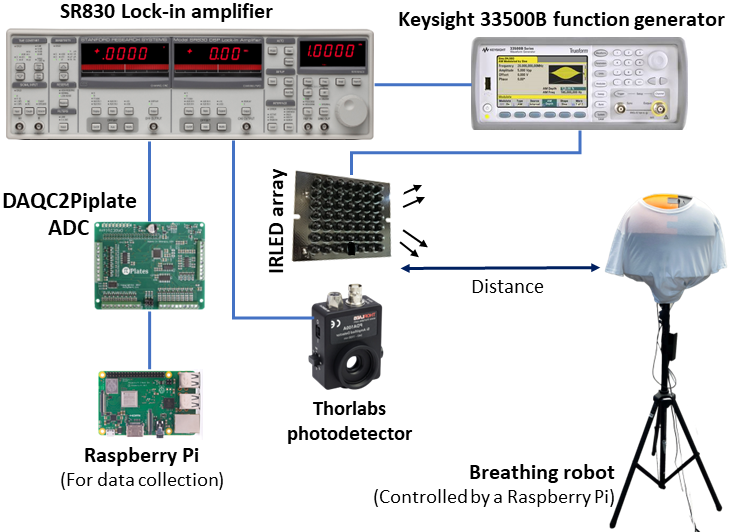}
 \caption{The experimental setup diagram of the LWS system used for data collection.}
 \label{fig:setup}
 \end{figure}

\subsection{LWS Hardware Setup}
\begin{figure}[t]
 \centering
 \includegraphics[width=0.36\textwidth]{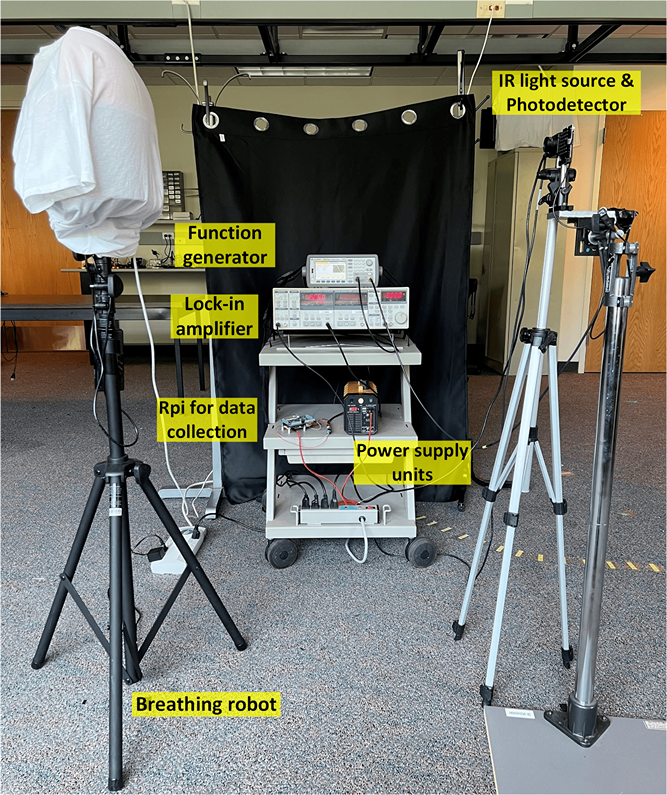}
 \caption{The experimental setup showing the key components of the LWS system used for data collection.}
 \label{fig:setup2}
 \end{figure}

A light-wave sensing system was developed for collecting respiration data using infrared light. The system utilized an IRLED matrix source (FY-48 940 nm IR lamp board~\cite{transmitter}), consisting of 48 IRLEDs, as the light source. The highest light intensity was achieved when the source was connected to $\approx$12\,V. The light source was modulated by a 1\,kHz sinusoidal voltage wave of 3.8\,V peak-to-peak amplitude and 8.1\,V DC offset from a Keysight 33500B function generator~\cite{funcgen}. These voltage settings were chosen to keep the light intensity high enough (by applying 11.9\,V peak voltage), while maintaining linear LED operation. For collecting the reflected light from the chest of the breathing robot, Thorlabs PDA100A photodetector~\cite{detector2} with a converging lens of 25.4\,mm focal length was used. The photodetector consisted of a p-i-n photodiode and a transimpedance amplifier with adjustable gain. For lock-in detection, SR830DSP frequency lock-in amplifier~\cite{lockin1} was utilized. Finally, for data collection and storage,  a second Raspberry Pi preceded by an analog-to-digital converter (DAQC2Piplate ADC~\cite{adc}) was used. Fig.~\ref{fig:setup} illustrates a diagram of the setup used for generating and collecting breathing data, while Fig.~\ref{fig:setup2} presents a comprehensive image of the overall setup, highlighting its core components as described earlier.

\subsection{Data Collection}
\begin{figure}[t]
 \centering
 \includegraphics[width=0.36\textwidth]{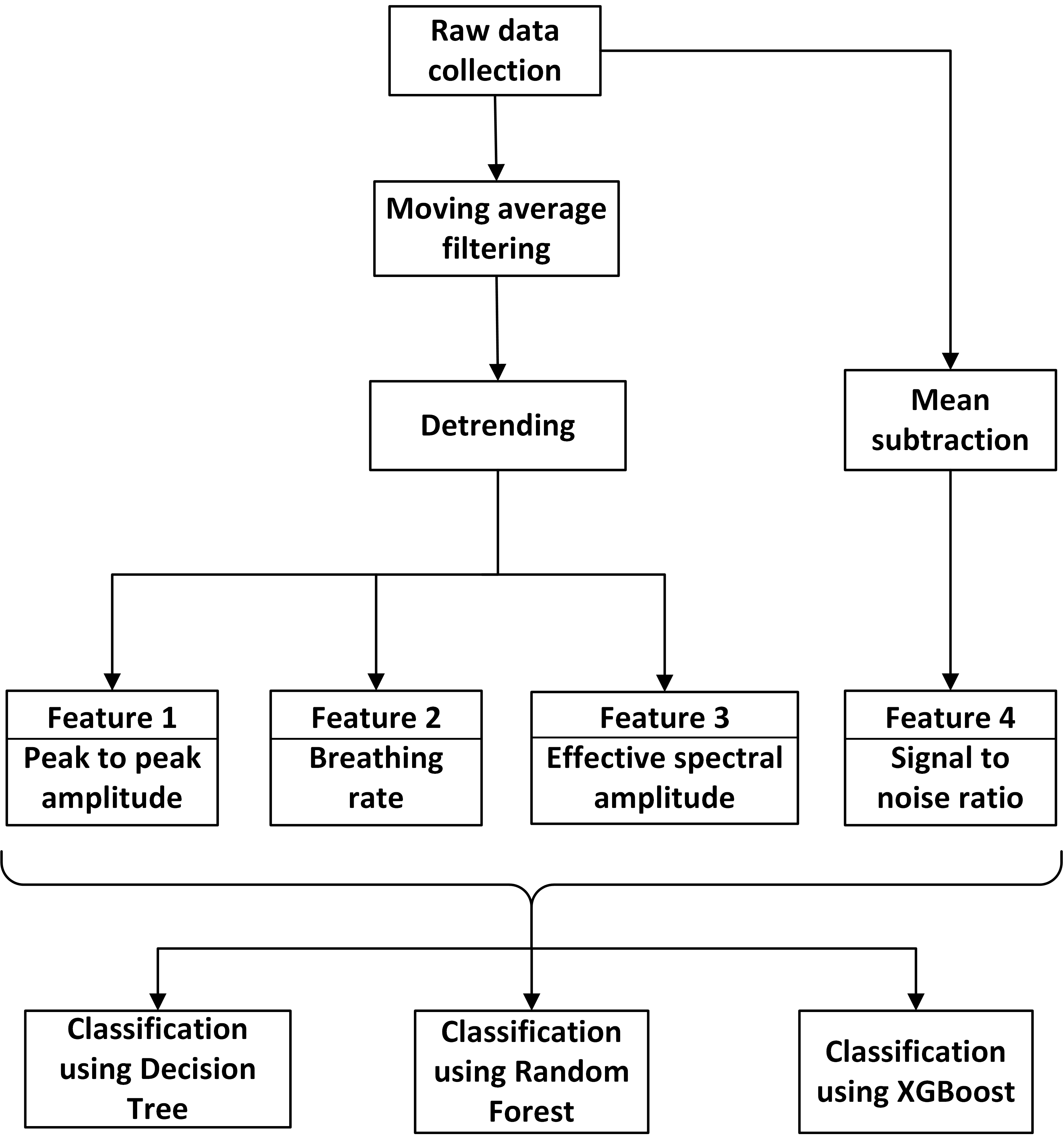}
 \caption{Flow diagram of the approach towards the breathing anomaly detection problem.}
 \label{fig:approach}
 \end{figure}

Breathing anomaly detection was conducted using machine learning-based classification of labeled breathing data collected using the developed IR sensing setup. The study utilized seven distinct breathing patterns as individual data classes, as described in Section~\ref{sec:human_breath_pattern}. To ensure data integrity, the system needed to determine the validity of the collected data. Certain factors like subject movement, proximity to others, excessive environmental noise, or system malfunction could lead to the collection of erroneous data which needed to be discarded or recollected. To detect such non-ideal conditions, a separate data class called `faulty data' was introduced. Therefore, the breathing anomaly detection problem was reduced to an eight-class classification task. The classes included eupnea, apnea, tachypnea, bradypnea, hyperpnea, hypopnea, Kussmaul's breathing and faulty data. Fig.~\ref{fig:approach} presents the comprehensive workflow for breathing anomaly detection, depicting key steps involved in data collection, pre-processing, feature extraction, and classification.

 \begin{table}[!ht]
\renewcommand{\arraystretch}{1.21}
\centering
\caption{Characteristics and number of collected data of each breathing class.}
\begin{center}
\begin{tabular}{|P{.65cm}|M{1.4cm}|P{1.42cm}|P{1.4cm}|P{1.77cm}|} 
 \hline
\bf{Class} & \bf{Class name} &  {\centering \bf{Breathing rate (BPM)}} &  {\centering \bf{Breathing depth (\%)}} & {\centering \bf{Number of data instances}}\\
 \hline
  \bf{0} & Eupnea & 12-20 & 30-58 & 300\\
  \bf{1} & Apnea & 0 & 0 & 300\\
  \bf{2} & Tachypnea & 21-50 & 30-58 & 300\\
  \bf{3} & Bradypnea & 1-11 & 30-58 & 300\\
   \bf{4} & Hyperpnea & 12-20 & 59-100 & 300\\
   \bf{5} &Hypopnea & 12-20 & 1-29 & 300\\
   \bf{6} &Kussmaul's & 21-50 & 59-100 & 300\\
   \bf{7} &Faulty data &Any &Any & 300\\
     \hline
      \multicolumn{4}{|r|}{\RaggedRight \bf{Total}} & 2400\\
 \hline
\end{tabular}
\end{center}
\label{table:classes}
\end{table}

\begin{figure}[!htbp]
    \centering
    \includegraphics[width=.505\textwidth,height=.38\textwidth]{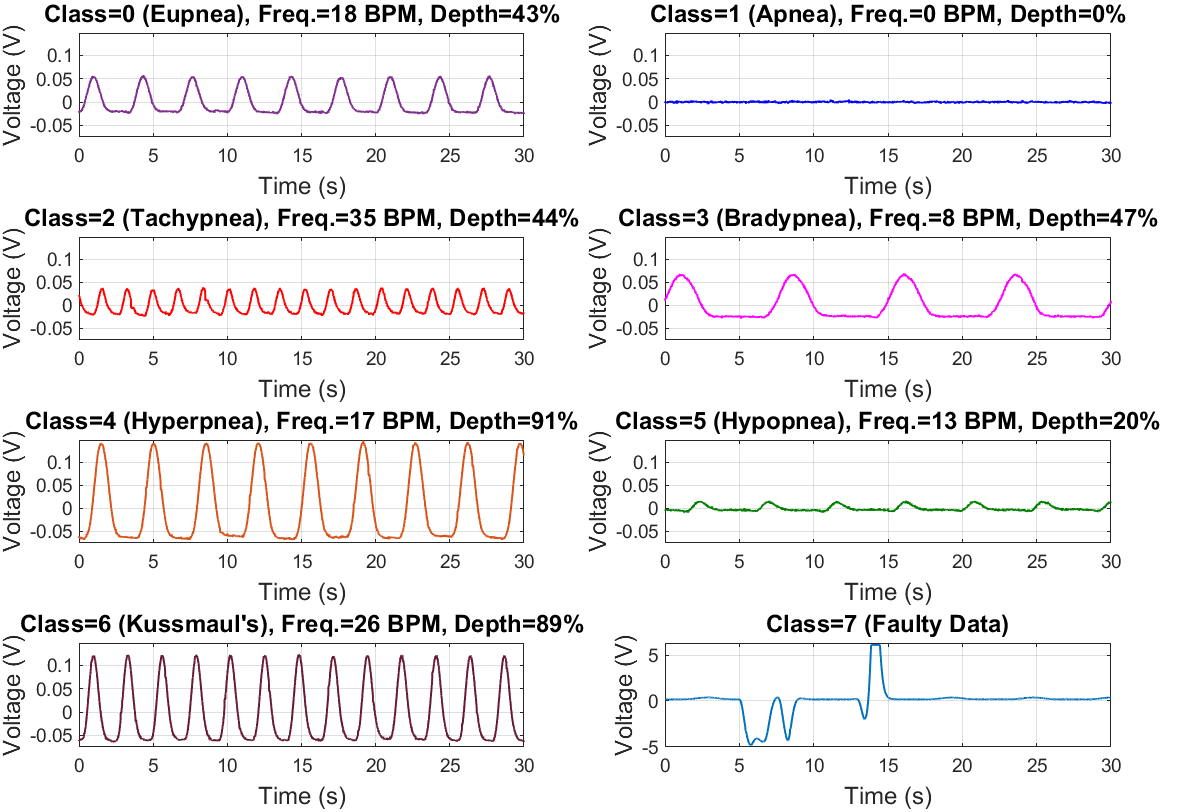}
    \caption{Time domain representation of sample data from each class (the distance between the source-photodetector and the robot was 1\,m).}
    \label{fig:data_visual}
\end{figure}

As the first step of breathing anomaly detection plan, data were collected using the IR light-wave sensing setup and the robot. $\sin^6$ pattern was used for generating breathing data because of its visual similarity to real human breathing. 
The time constant at the lock-in amplifier was configured to 100\,ms. To prevent voltage saturation during data collection, the gain at the photodetector was set to 40\,dB, which was halfway across the full span. The sensitivity of the lock-in amplifier was carefully adjusted to maintain the resting voltage level near the middle of its voltage range, allowing for maximum voltage swing without saturation. The breathing offset, representing the initial position of the robot's chest, was set to zero. Data collection primarily took place at three different distances (0.5\,m, 1\,m and 1.5\,m) between the photodetector and the robot during the daytime. The windows of the room were unshaded and internal lighting was common for an office environment during data collection. The chosen ranges of breathing rate and depth for each class, as discussed in Section~\ref{sec:human_breath_pattern}, along with the number of collected data instances, are summarized in Table~\ref{table:classes}. To ensure class uniformity and enhance generalization capability, an equal number of 100 data instances were recorded for each class at every chosen distance. To create faulty data for class 7, various intentional disturbances were introduced, including walking closely around the setup, manually interrupting the line of sight, and intentionally causing system malfunctions intermittently during data collection. Each data and corresponding timestamps were collected for 30\,s duration and stored in the Raspberry Pi. The sampling frequency used was 100\,Hz which provided more than sufficient resolution to capture even the fastest breathing frequency, 50\,BPM or 0.8\,Hz as per Table~\ref{table:classes}. Later, a MATLAB script was used to save the voltage amplitudes in a single CSV (Comma Separated Value) file along with their corresponding class labels. In that CSV file, each row denoted one data instance that contained $30 \text{\,s}\times100 \text{\,Hz} = 3000$  data points and one integer between 0 to 7 (inclusive) for denoting its class label. Fig.~\ref{fig:data_visual} shows several samples of raw data, one from each class, collected at a distance of 1\,m after subtracting mean from the data. For better comparison among data classes,  the first 7 graphs were plotted with the same voltage range on the y-axis. However, the graph representing faulty data required a larger span in y-axis to display the complete signal.

\begin{figure}[!htbp]
    \centering
    \includegraphics[width=.46\textwidth]{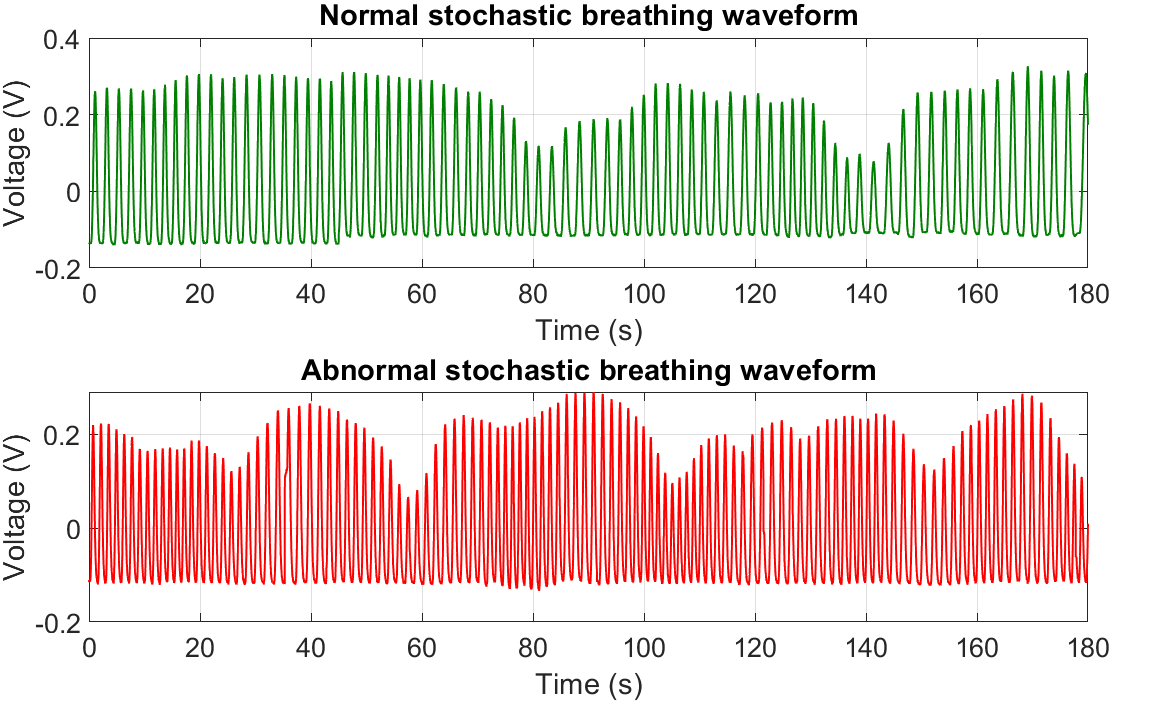}
    \caption{Realistic (stochastic) respiration data of normal and abnormal classes, collected at 1\,m distance.}
    \label{fig:stochplot}
\end{figure}

In real life applications, human respiration frequency and depth may exhibit stochastic drift within distinct ranges for normal and abnormal breathing. Additionally, the subject may be located beyond 1.5\,m, especially in case of health monitoring at home. To assess the feasibility of the proposed system in such realistic situations, measurements of stochastic respiration waveforms were conducted over one-hour intervals. The breathing phantom was placed at distances ranging from 0.5\,m to 2.5\,m in 0.5\,m increments. At each distance, two classes of breathing were measured. The first respiration pattern was designed with a rate distribution typical of a healthy adult. The second pattern featured traits like elevated respiration rates of increased standard deviation, which are often linked to unhealthy breathing. The mean and variance of breathing frequency for the two patterns were selected in accordance with known realistic values~\cite{acute}. Both classes incorporated amplitude modulation with a minimum amplitude of 0.3\,V, ensuring the inclusion of periods of both shallow and deep breathing in each pattern. Over the course of an hour, more than a thousand respiratory cycles occur, covering a diverse ensemble of respiration rates and amplitudes. The collected stochastic respiration waveforms were segmented at each 60\,s intervals for the purpose of machine learning-based anomaly detection. 50 data instances were taken from each of 5 considered distances for each class. Hence, there were $5\times50=250$ data samples per class and $250+250=500$ data instances in total, with normal and abnormal respiration classes marked by class 0 and class 1. Normal and abnormal stochastic respiration data for 3 minutes for each class, collected at 1\,m distance, have been visualized in Fig.~\ref{fig:stochplot}. It is essential to note that respiratory rates and depths can vary even within a single instance of these data. The data lengths were kept longer than those of previous no-drift scenarios to preserve and emphasize their stochastic characteristics.

\subsection{Data Processing}
The collected data was affected by noises from the environment and the setup, which was subsequently filtered using a $k$-points moving averaging technique. Through trial and error, it was determined that a 50-points moving averaging provided the best results for all data, effectively smoothing the data while preserving the sinusoidal variation associated with different breathing frequencies. Some data had a drift or trend in them which added a DC component to the data and increased the peak-to-peak amplitude. Thus, those data were prone to underestimating breathing rate and overestimating breathing depth. In order to get around this problem, the data was detrended by fitting a polynomial of order 5 to the data and subtracting that polynomial trend from the data. Detrending successfully centered the data around 0\,V and eliminated most of the drift.

\begin{figure}[t]
    \centering
    \includegraphics[width=.42\textwidth]{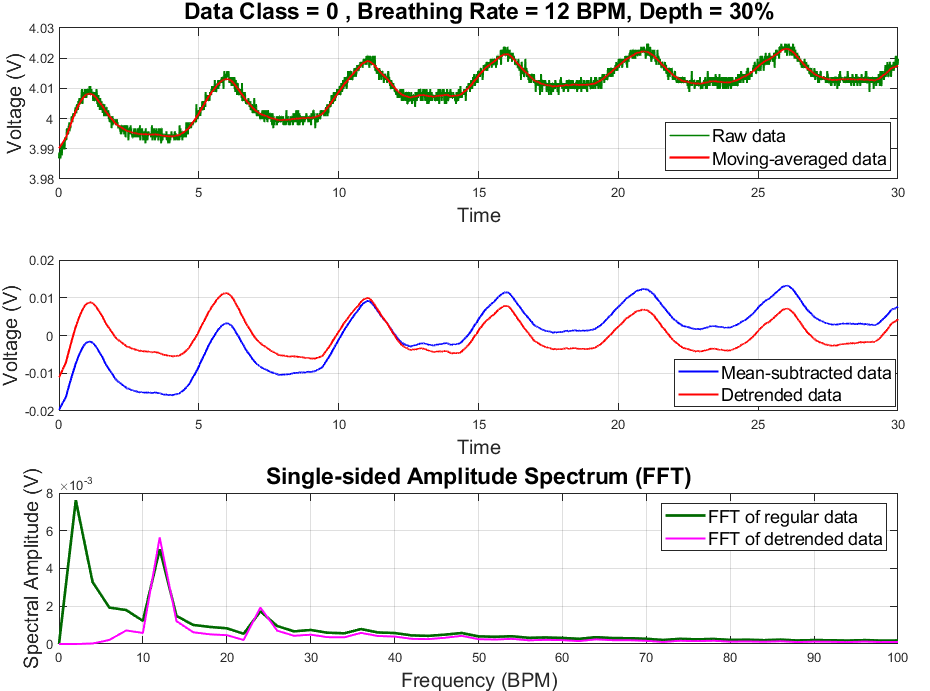}
    \caption{Time domain and frequency domain plots of sample raw and processed data from class 0 (Eupnea) at 1.5\,m distance.}
    \label{fig:one_data_plot}
\end{figure}

Fig.~\ref{fig:one_data_plot} shows an instance of data from class 0 collected at a distance of 1.5\,m along with its processed versions and frequency domain representation. Due to greater distance, the raw data in the first subplot (green colored) exhibited more noise compared to the samples in Fig.~\ref{fig:data_visual}. Also, this data exhibited a noticeable upward trend. In the second subplot, the effect of detrending became clear when the detrended data (red colored) was compared with the mean-subtracted version (blue colored) of the filtered data. The peak-to-peak amplitude of this data without detrending was 0.033\,V, exceeding the actual peak-to-peak amplitude of the underlying periodic variation. After detrending, a more accurate estimation of 0.022\,V was obtained. In the third subplot, the highest spectral amplitude was found at 2\,BPM without detrending which was erroneous. The actual breathing frequency was at the secondary peak in the spectrum occurring at 12\,BPM. Detrending improved the breathing frequency estimation by suppressing the false peaks at low frequencies.

\section{Feature extraction}
\label{sec:feature_ex}
Feature extraction is an important step in machine learning-based data classification. After detrending, four handcrafted features were extracted from the collected data using MATLAB code for the following three cases:

\begin{enumerate}
\item Single distance at a time or at 0.5\,m, 1\,m and 1.5\,m independently.
\item Two consecutive distances at a time i.e., 0.5\,m and 1\,m together and also 1\,m and 1.5\,m together.
\item All three distances together.
\end{enumerate}

The features for each data were saved in separate rows in CSV files along with the class label for each row. Thus, labeled features were prepared for the subsequent classification task. The details of extracted handcrafted features are provided as follows.

\subsubsection{Peak-to-peak Amplitude}
\label{subsubsec:p2p_ampl}
Peak-to-peak amplitude is defined as the difference between the maximum and minimum values of the processed data. It roughly represents a number proportional to the breathing depth in each class (except the faulty data class), when the test distance remains constant. When the test distance is changed, the value of this feature is shifted to a different range, even for data belonging to the same class, because of changed levels of received light intensity and noise. 

\subsubsection{Breathing Rate}
\label{subsubsec:rate}
Breathing rate is the frequency with the highest spectral amplitude in the frequency domain representation of the signal (again, it is not applicable in apnea and faulty data classes). 
Discrete Fourier Transform (DFT) of the time domain data $z[n]$ is taken using
\begin{equation}
   \label{eq:FFT1}
                 Z[k]=\sum_{n=0}^{L-1} z[n]e^{\frac{-j2\pi kn}{L}},
  \end{equation}
where L is the length of the dataset and its DFT. $Z[k]$ contains different frequency components present in the signal $z[n]$. The frequency index with the maximum magnitude is found to be $k_{\max} = \underset{k}{\text{arg max}}~|Z[k]|$. Then, the corresponding frequency or the breathing rate, $f_{max}$, is calculated using
\begin{align}
        f_{max} = \frac{k_{max}f_s}{L} \text{\,Hz} = \frac{60k_{max}f_s}{L} \text{\,BPM},
\end{align}
where $f_s$ is the sampling frequency used to collect data. Fast Fourier Transform (FFT) algorithm was used to find DFT of the processed data in order to extract breathing rate. Breathing rate stays in the same range for data collected at all distances. In contrast, error in breathing rate estimation increases with increasing distance because of increasing noise level.

\begin{figure}[!htbp]
    \centering
    \includegraphics[width=.44\textwidth]{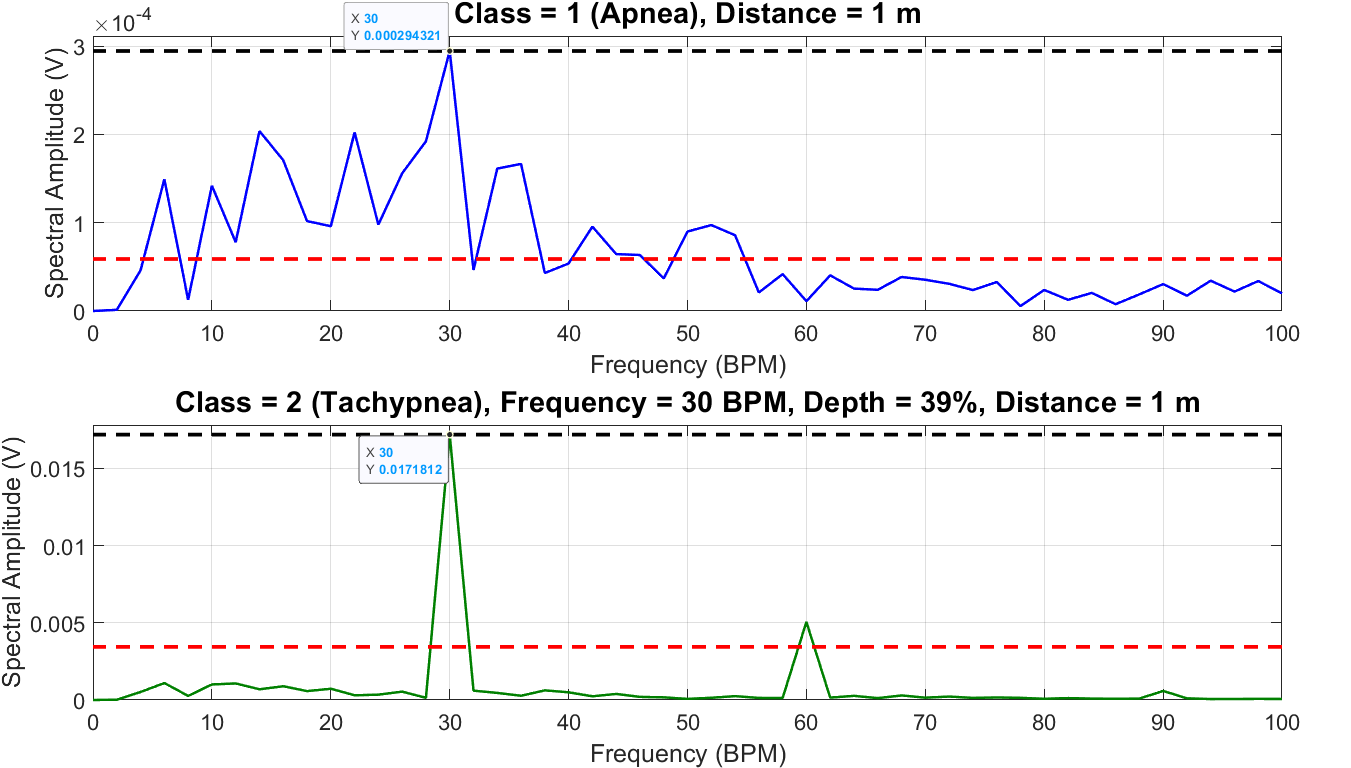}
    \caption{Frequency domain plots of sample processed data from class 1 and 2, to show the calculation process and usefulness of effective spectral amplitude feature.}
    \label{fig:eff_spec_amp2}
\end{figure}

\subsubsection{Effective Spectral Amplitude}
\label{subsubsec:eff_spec_ampl}
It is the percentage count of frequencies in the spectrum of data after detrending whose spectral amplitude is greater than or equal to a predefined threshold value. The threshold is determined as a percentage of peak spectral amplitude. If the peak magnitude in the data spectrum is $H$, $t$ is the percentage threshold, and A is the spectrum vector, then effective spectral amplitude
\begin{equation}
    S=\frac{l(B)}{l(A)}\times100\%, ~B = \{x \in A |x \ge tH\},
    \label{eq:eff_spec_amp}
\end{equation}
where $l()$ represents the length of its parameter vector. This number helps differentiate between a very clean spectrum with just a few large peaks (which are the breathing rate and its harmonics) and a spectrum with multiple peaks of similar magnitude at different frequencies. The threshold $t$ was chosen to be 20\% and only first 100 frequency points in the spectrum were considered ($l(A)=100$ in equation~(\ref{eq:eff_spec_amp})), because the rest of the frequencies had negligible contributions to the signal. 

As an example, in the first subplot of Fig.~\ref{fig:eff_spec_amp2}, an instance of data from the apnea class was found to have its highest peak occurring at 30\,BPM, which is an incidental occurrence. It has many other peaks of magnitudes greater than the threshold (marked by red line). Hence, it is not indicative of a breathing rate of 30\,BPM, as evident from its larger effective spectral amplitude ($S=20$). On the other hand, an actual breathing data of tachypnea class with breathing frequency 30\,BPM has a cleaner peak at 30\,BPM in its spectrum with $S = 2$, as shown in the second subplot of the same figure. The threshold is assumed to be 20\% of the peak magnitude in both cases.

\subsubsection{Signal-to-noise Ratio}
\label{subsubsec:SNR}
The average signal-to-noise ratio (SNR) of a signal is an indicator of signal quality and is defined as the ratio of average signal power and average noise power. The photodetector was a power detector that detected the signal power and noise power together. If $z[n]$ is the mean-subtracted data received by the photodetector at a particular distance and $x_{noise}[n]$ is the mean-subtracted noise data at that distance, then SNR in decibel can approximately be defined by 
\begin{equation}
    \text{SNR}_{dB}=20\log_{10}{\left(\frac{\sqrt{\sum\limits_{n=1}^{N}|z[n]|^2}}{\sqrt{\sum\limits_{n=1}^{N}|x_{noise}[n]|^2}}\right)},
    \label{eq:snr}
\end{equation}
considering root-mean-square averages of the signals. Here, both of $z[n]$ and $x_{noise}[n]$ has the same length $N$. SNR value will decrease with increasing distance in the same data class because of increasing level of noise. Also, SNR will decrease when the breathing depth is decreased, as in class 3 (bradypnea), because of lower signal power.

The first three features did not need the distance information to be calculated, hence they were calculated once and used in all three cases above as needed. But the last feature, SNR, was dependent on noise power which varied with distance. Noise data for 30\,s were collected by the IR sensing setup at the chosen three distances by keeping the robot stationary (not breathing). When data collected at only one distance was considered for classification, noise data of that distance was used to calculate SNR. For a mixed set of data (either 2 or 3 distances together), an average noise was considered.

\section{Data Classification}
\label{sec:data_classification}
In this study, three machine learning algorithms were employed for data classification in the context of anomaly detection. The algorithms were decision tree, random forest, and XGBoost. The data classes, as defined in Table II, were differentiated by two categorical variables, namely breathing rate and breathing depth, which can be effectively represented by a decision tree-like structure. Hence, decision tree model did the classification better than other available machine learning models with the breathing dataset. Random forest and XGBoost are ensembles of multiple decision trees, and therefore, were expected to perform better than a single tree. The feature vectors extracted from the data were utilized to train these machine learning models using Scikit-learn library in Python. In order to see the effect of distance on classification accuracy, classification was initially done separately on data collected at one particular distance at a time. However, in real-world scenarios, users may not consistently be at a fixed distance. To account for this variability, mixed sets of data collected at different distances were subsequently employed for feature extraction and classification. Each time, the feature set for the data was randomly split into training and test sets, with 80\% feature data allocated for training sets, and 20\% for testing. The models were trained using the training dataset, and their performance was assessed by calculating test accuracies. 10-fold cross-validation was performed on the dataset and models were evaluated using average accuracy scores and confusion matrices. The three machine learning algorithms used are briefly explained as follows.

\subsubsection{Decision Tree}
\label{subsubsec:dt}
Decision tree is one of the most effective non-parametric supervised machine learning models used for classification, prediction and data mining. This model starts the classification process by partitioning the dataset into two or more mutually exclusive subsets based on the values of a root node, then continues partitioning through internal nodes and ends at leaf nodes. Thus, it forms a tree-shaped graph overall. 

\subsubsection{Random Forest}
\label{subsubsec:rf}
Random forest is an ensemble version of decision tree  model. It makes use of Bootstrap Aggregating or Bagging technique~\cite{Lee2020} in the following two ways to build multiple uncorrelated decision trees and finally takes majority vote to predict the class labels:

\begin{enumerate}[label=(\roman*)]
\item If the dataset is $x_1, x_2,...,x_n$, then random forest takes random samples of size $n$ with replacement from this dataset $m$ times and builds $m$ different decision trees with them parallelly.
\item If the features defining the dataset are $f_1, f_2,..., f_p$, then it considers a random subset of features of size $k$ ($k<p$) instead of all features, while splitting each node of decision trees to create new branches.
\end{enumerate}

\begin{table}[!ht]
\renewcommand{\arraystretch}{1.21}
\centering
\caption{Classification accuracies with 10-fold cross-validation.}
\begin{center}
\begin{tabular}{|M{2.6cm}|P{1.3cm}|P{1.3cm}|P{1.3cm}|} 
\hline
\multirow{2}{*}{\bf{Distances}} & \multicolumn{3}{p{4.9cm}|}{\centering\bf{Average test accuracy scores with cross-validation}}\\
\cline{2-4}
& \bf{Decision tree} & \bf{Random forest} & \bf{XGBoost}\\
  \hline
\bf{0.5\,m}  & 95.88\%   & 96.75\%    & 96.25\%\\
\bf{1\,m}  & 94.75\%    & 96.38\% &  95.75\% \\
\bf{1.5\,m}  & 88.13\%   & 90.88\% &  90.63\%  \\
\bf{0.5\,m and 1\,m}  & 91.25\%    & 93.19\% &  92.75\%\\
\bf{1\,m and 1.5\,m}  & 83.38\%   & 86.88\% &  86.88\%\\
\bf{0.5\,m, 1\,m and 1.5\,m}  & 84.54\%   & 87.37\% & 86.96\%\\
  \hline
\end{tabular}
\end{center}
\label{table:accuracy_with_cv}
\end{table}

\begin{figure}[!htbp]
    \centering
    \includegraphics[width=.47\textwidth]{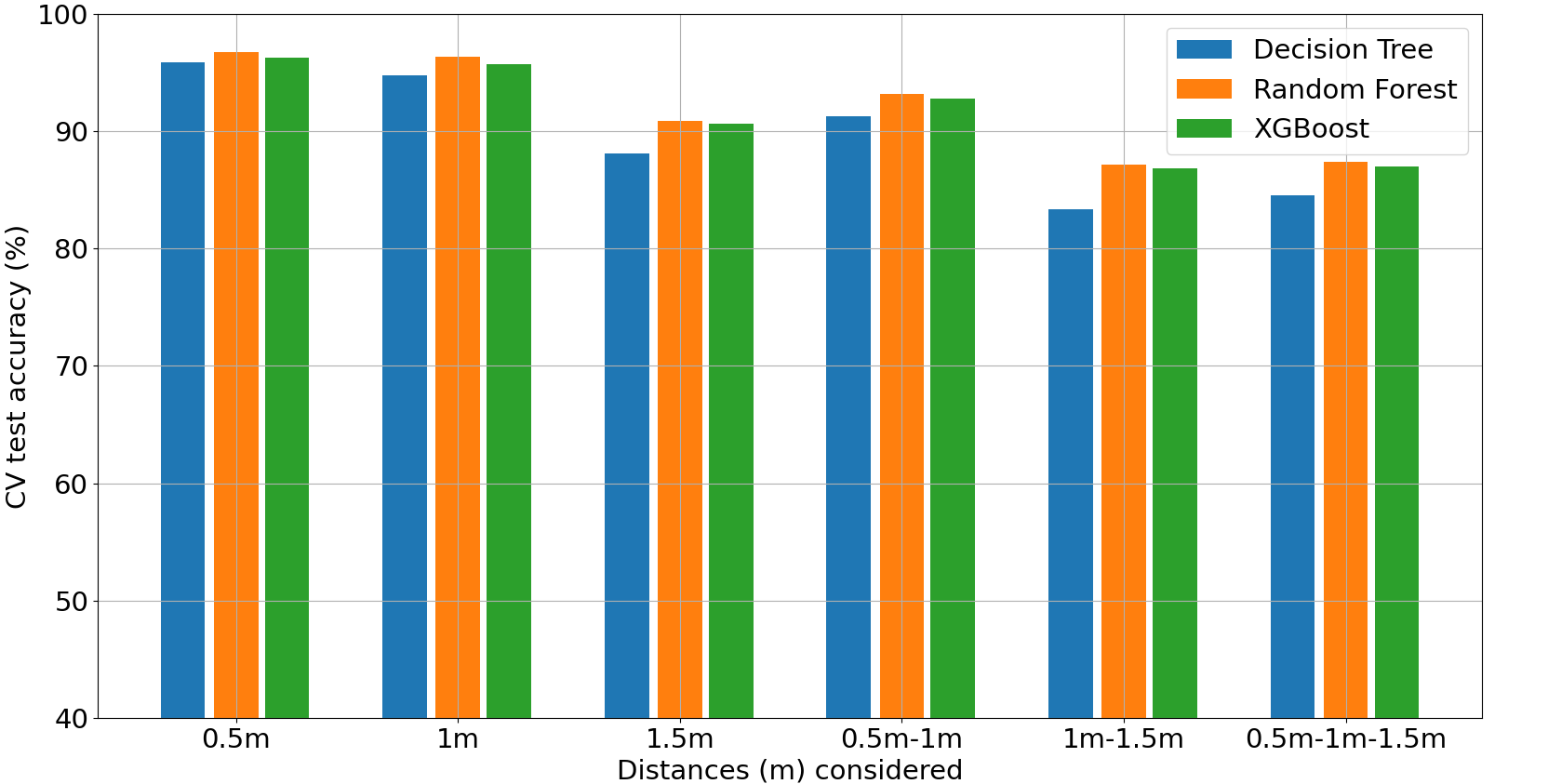}
    \caption{Cross-validation based test accuracy plots for decision tree, random forest and XGBoost models for different distances.}
    \label{fig:acc_bar_chart}
\end{figure}

\begin{figure*}[!hbt] \centering
\begin{subfigure}[!hbt]{0.28\textwidth}
 \centering
\includegraphics[width=\textwidth]{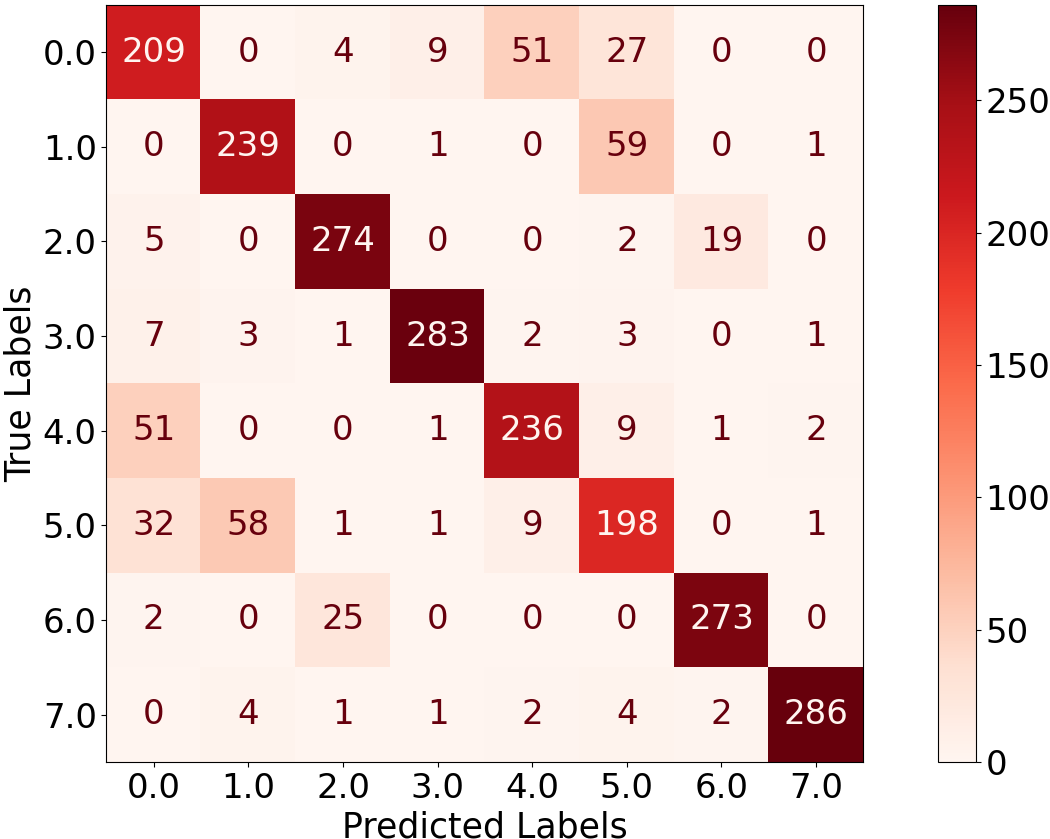}
\caption{}
\label{fig:10_dt_cm_all}
\end{subfigure}
\hfill
\begin{subfigure}[!hbt]{0.28\textwidth}
 \centering
\includegraphics[width=\textwidth]{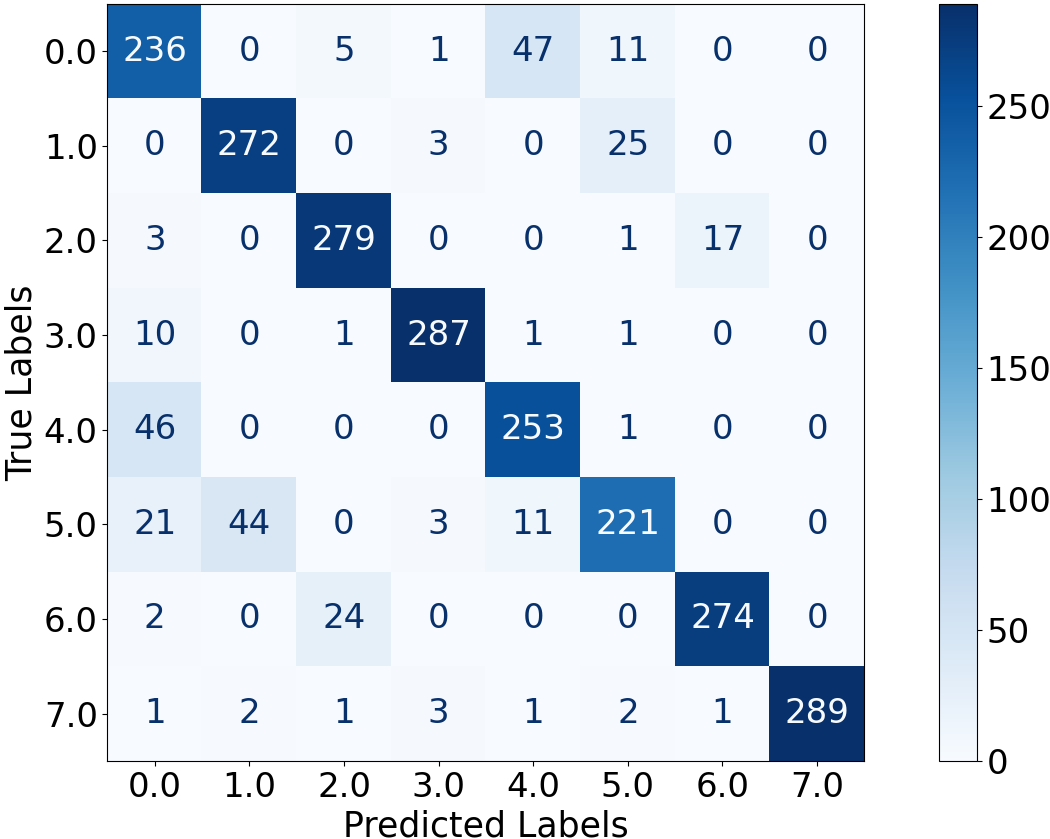}
\caption{}
\label{fig:11_rf_cm_all}
\end{subfigure}
\hfill
\begin{subfigure}[!hbt]{0.28\textwidth}
 \centering
\includegraphics[width=\textwidth]{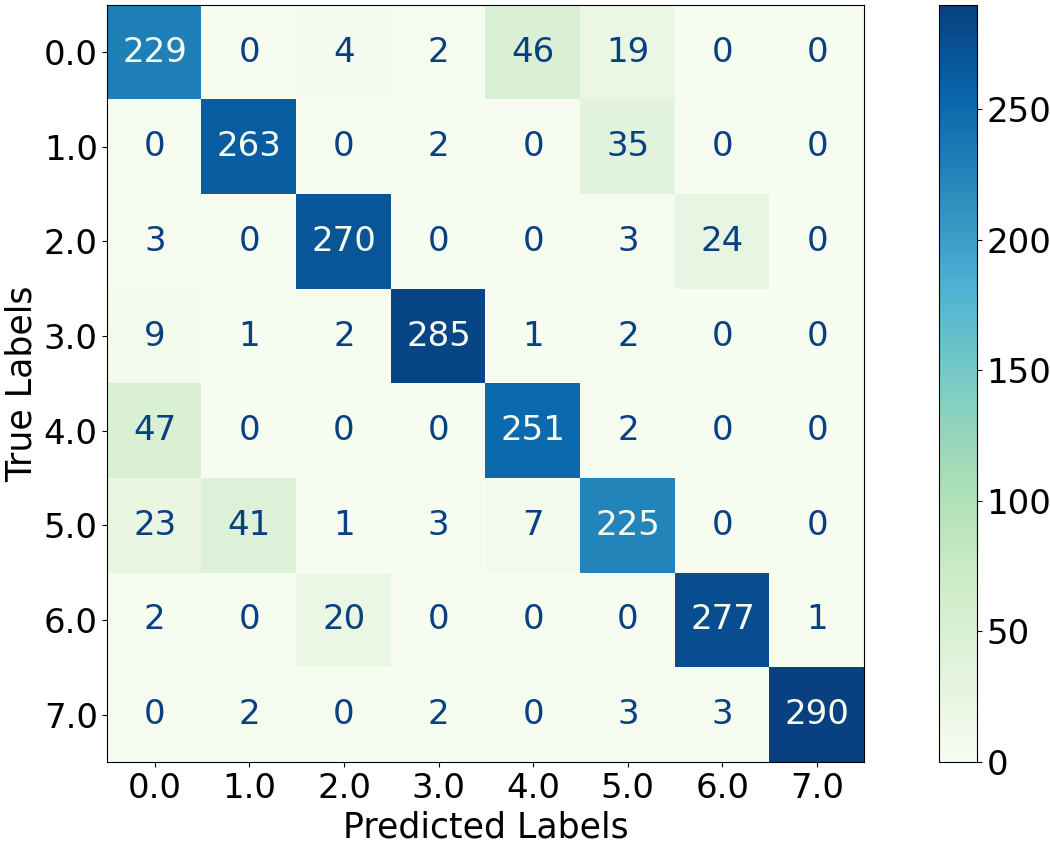}
\caption{}
\end{subfigure}
\hfill
\caption{Confusion matrices for 8-class classification  of breathing data considering all distances (0.5\,m, 1\,m and 1.5\,m) together found through cross-validation with (a) decision tree, (b) random forest and (c) XGBoost algorithms.}
\label{fig:cm1}
\end{figure*}

\begin{figure*}[!hbt] \centering
\begin{subfigure}[!hbt]{0.43\textwidth}
 \centering
\includegraphics[width=\textwidth]{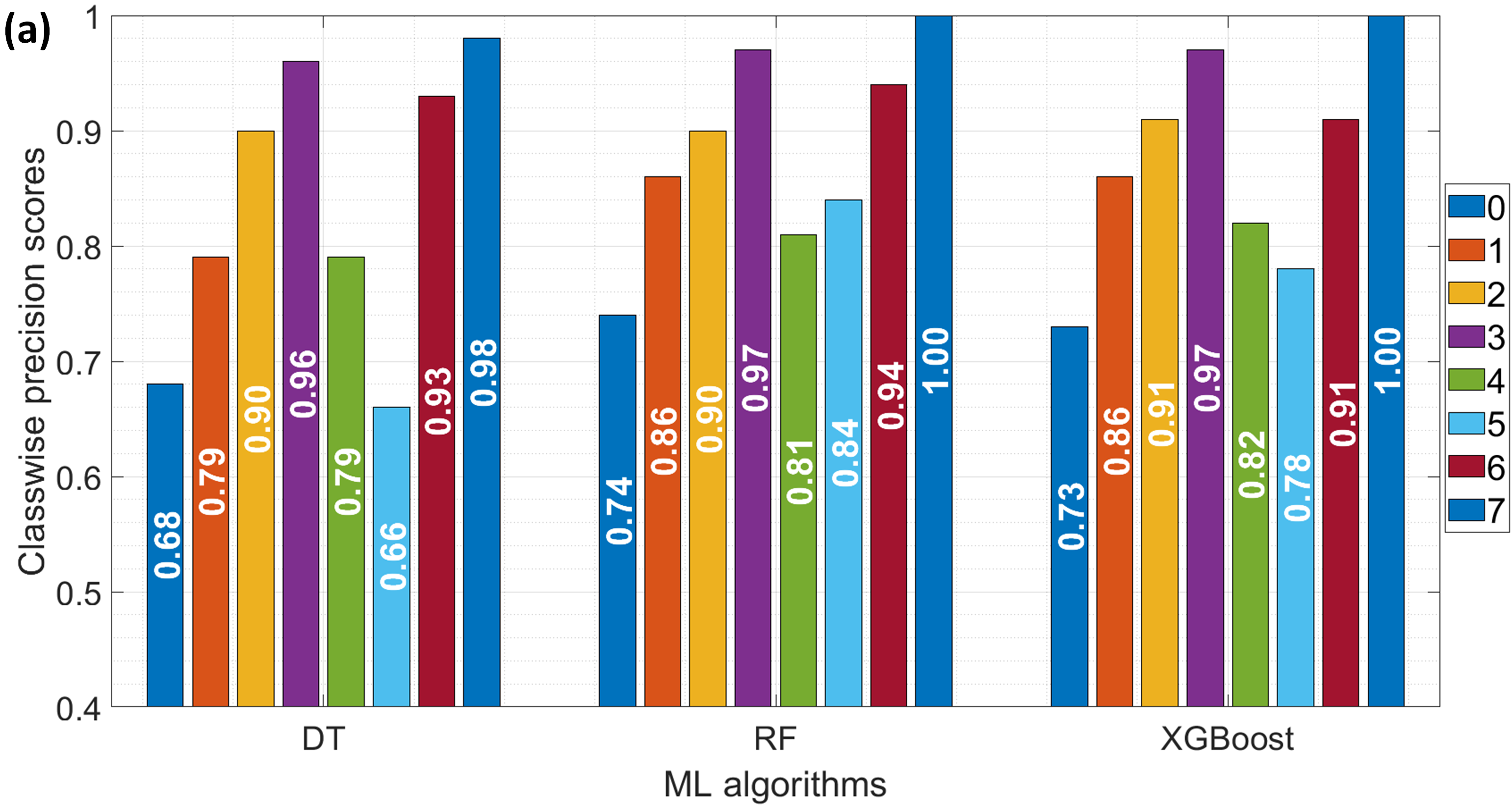}
\label{fig:prec}
\end{subfigure}
\hspace{6mm}
\begin{subfigure}[!hbt]{0.43\textwidth}
 \centering
\includegraphics[width=\textwidth]{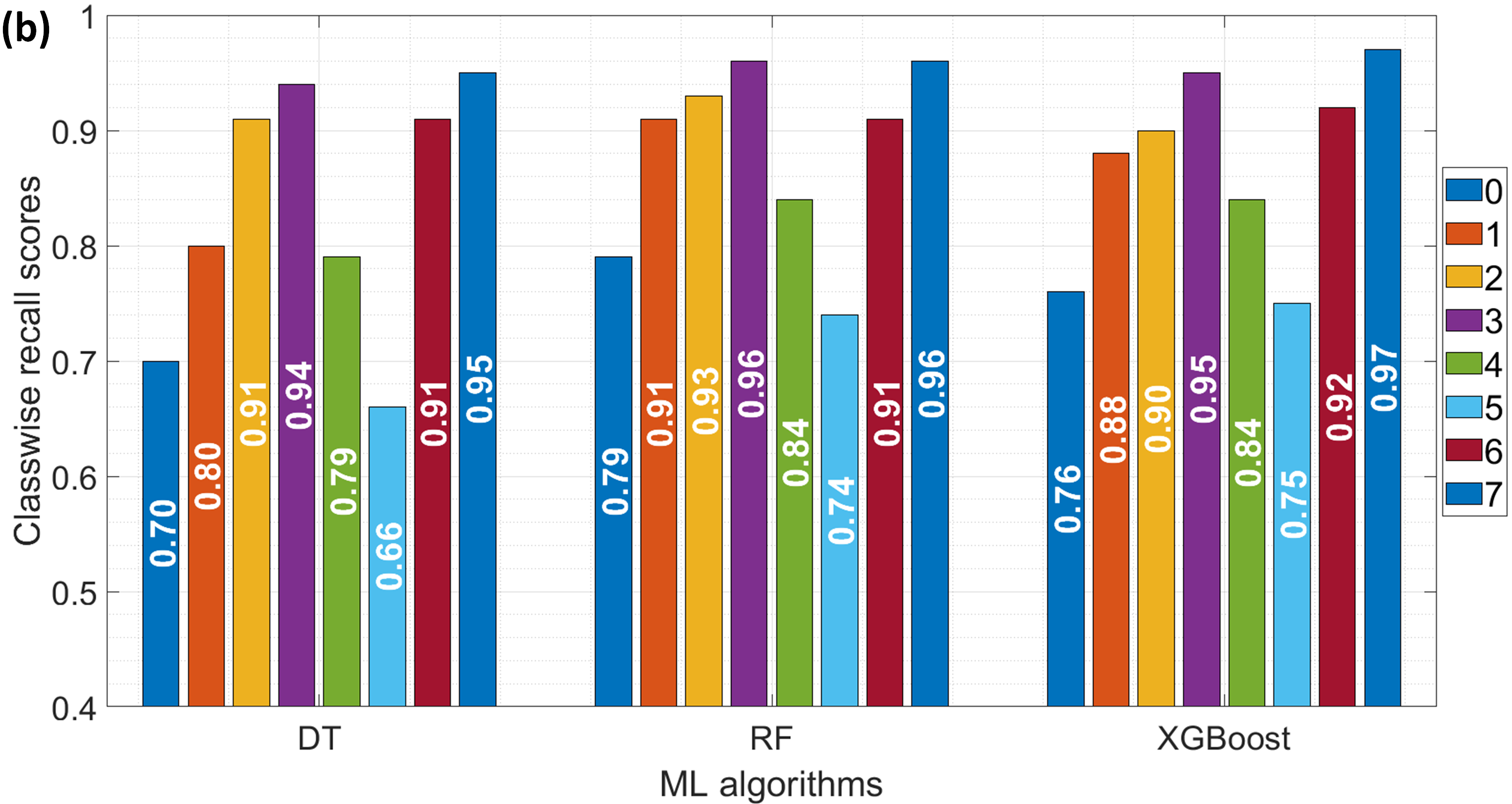}
\label{fig:rec}
\end{subfigure}
\caption{Classwise (a) precision and (b) recall plots for 8-class classification using decision tree, random forest and XGBoost models considering all distances (0.5\,m, 1\,m and 1.5\,m) together.}
 \label{fig:prec_rec_bar_chart}
\end{figure*}

\subsubsection{XGBoost}
\label{subsubsec:xg}
XGBoost is another widely used ensemble learning algorithm based on decision trees and the gradient boosting framework~\cite{sagi2018ensemble}. It constructs a robust model by iteratively combining multiple weak models. Each tree is trained on a different subset of the data, learning from errors of previous trees. This mitigates overfitting and underfitting and enhances prediction accuracy. The algorithm assigns weights to trees during training, focusing on the most informative features and emphasizing trees contributing significantly to the final prediction.

\section{Model Evaluation}
\label{sec:model_eval}

\subsection{Results}
In this study, we evaluated the performance of the multiclass classification performed for breathing anomaly detection using some commonly used evaluation metrics, including $K$-fold cross-validation, confusion matrix, accuracy, precision and recall. Average accuracies achieved with cross-validation are reported in Table~\ref{table:accuracy_with_cv}. In order to facilitate a more effective comparison of accuracies across different distances and models, Fig.~\ref{fig:acc_bar_chart} presents a bar chart that displays the test accuracies obtained through cross-validation. Fig.~\ref{fig:cm1} presents the confusion matrices obtained through cross-validation for each model. In order to evaluate the reliability of the classification, precision and recall scores for each class were derived from the confusion matrices and visually presented in Fig.~\ref{fig:prec_rec_bar_chart}.

The 8-class classification system, which includes six different types of abnormal breathing, offers a deeper understanding of abnormalities and is beneficial in clinical settings. However, in the context of in-home health monitoring, the ability to differentiate between normal and abnormal breathing alone can be of greater importance. To address this, we performed a simpler three-class classification, where normal breathing was labeled as class 0, abnormal breathing as class 1, and faulty data as class 2. Class 0 and 2 consisted of all available data, with 300 samples each. To avoid class imbalance, 300 samples were randomly selected from the available 1800 abnormal breathing samples for class 1.  Decision tree, random forest, and XGBoost algorithms were applied with cross-validation for the intended 3-class classification, generating confusion matrices depicted in Fig.~\ref{fig:3classcm}. The average test accuracies were 91.56\% with decision tree, 93.33\% with random forest, and 92.44\% with XGBoost algorithm. Precision and recall scores were plotted too in Fig.~\ref{fig:3clsbar}.

Subsequently, a binary classification was conducted on the collected stochastic respiration data to distinguish between normal and abnormal breathing classes. The employed machine learning algorithms produced confusion matrices, as illustrated in Fig.~\ref{fig:stochastic_cm}. Through cross-validation, the average test accuracies were found to be 96\% with the decision tree, 96.8\% with the random forest, and 96.4\% with the XGBoost algorithm.


\subsection{Discussion}

We will analyze the results achieved to gain further insight in this section. $K$-fold cross-validation involves iteratively assigning each data to either the training or test dataset, allowing for a more robust evaluation by averaging the results across multiple models. Hence, in this study, the test accuracies obtained through $K$-fold cross-validation with $K=10$, as shown in Fig.~\ref{fig:acc_bar_chart}, provided a more reliable measure of model performance on unseen test data compared to the accuracy scores obtained without cross-validation.

The results revealed that the classification accuracies decreased for all three algorithms as the distance increased, mainly due to higher path-loss and lower signal-to-noise ratio in the received signal. Notably, ensemble models such as random forest and XGBoost consistently outperformed single decision trees at all distances. This improvement in accuracy was particularly pronounced in the last four rows of Table~\ref{table:accuracy_with_cv}, where the noise and intra-class variances in the data were higher due to longer distances and the combination of data collected at multiple distances. Ensemble models achieved better predictions by aggregating results from multiple decision trees, thus compensating for individual wrong predictions. Hence, ensemble models yielded more generalized models for respiratory anomaly detection, which were expected to perform better on unseen test data compared to single models. Among the ensemble models, random forest exhibited superior accuracy with the highest accuracy of 96.75\% at 0.5\,m distance. However, the accuracies obtained with XGBoost were also sufficiently good and closely aligned with those obtained with the random forest model. The last row in Table~\ref{table:accuracy_with_cv} is of particular importance because it represents the most realistic scenario through combining data from multiple distances. Hence, further analysis of the results is provided on this case only.

\begin{figure}[!hbt] \centering
\begin{subfigure}[!hbt]{0.155\textwidth}
 \centering
\includegraphics[width=\textwidth]{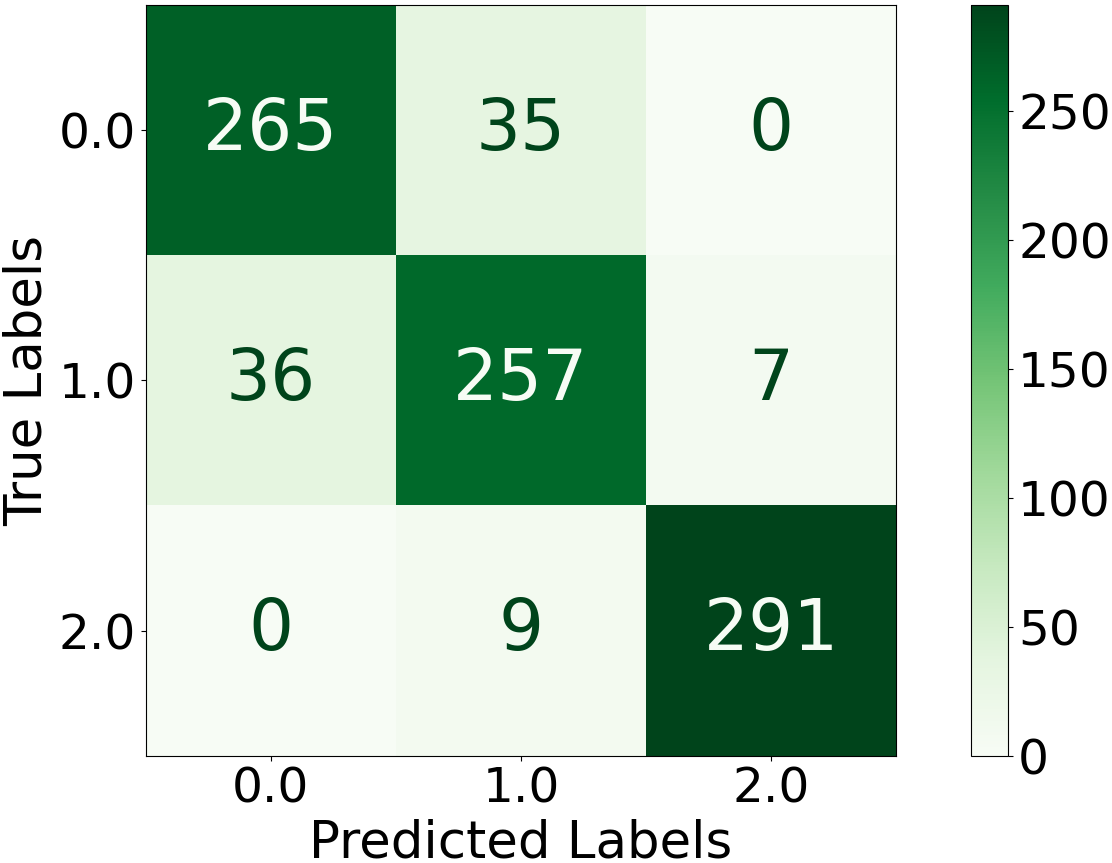}
\caption{}
\label{fig:1_dt_cm_all_3cls}
\end{subfigure}
\hfill
\begin{subfigure}[!hbt]{0.155\textwidth}
 \centering
\includegraphics[width=\textwidth]{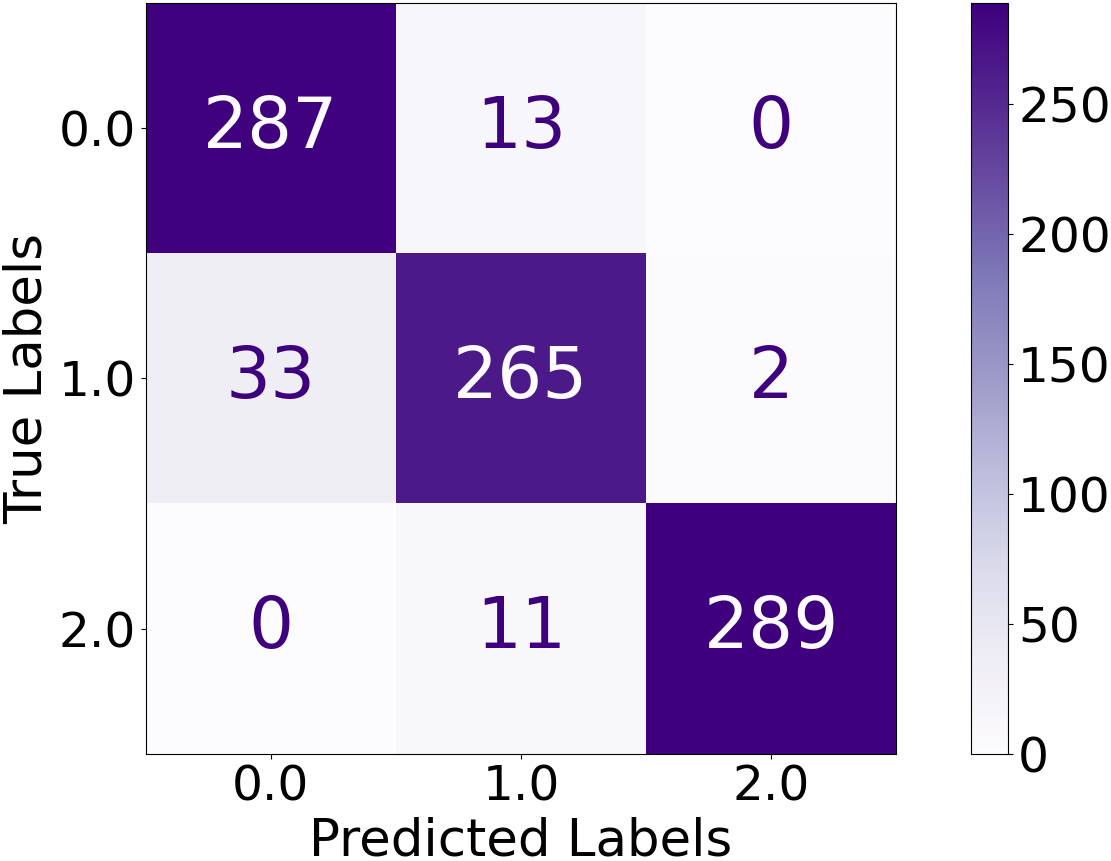}
\caption{}
\label{fig:2_rf_cm_all_3cls}
\end{subfigure}
\hfill
\begin{subfigure}[!hbt]{0.155\textwidth}
 \centering
\includegraphics[width=\textwidth]{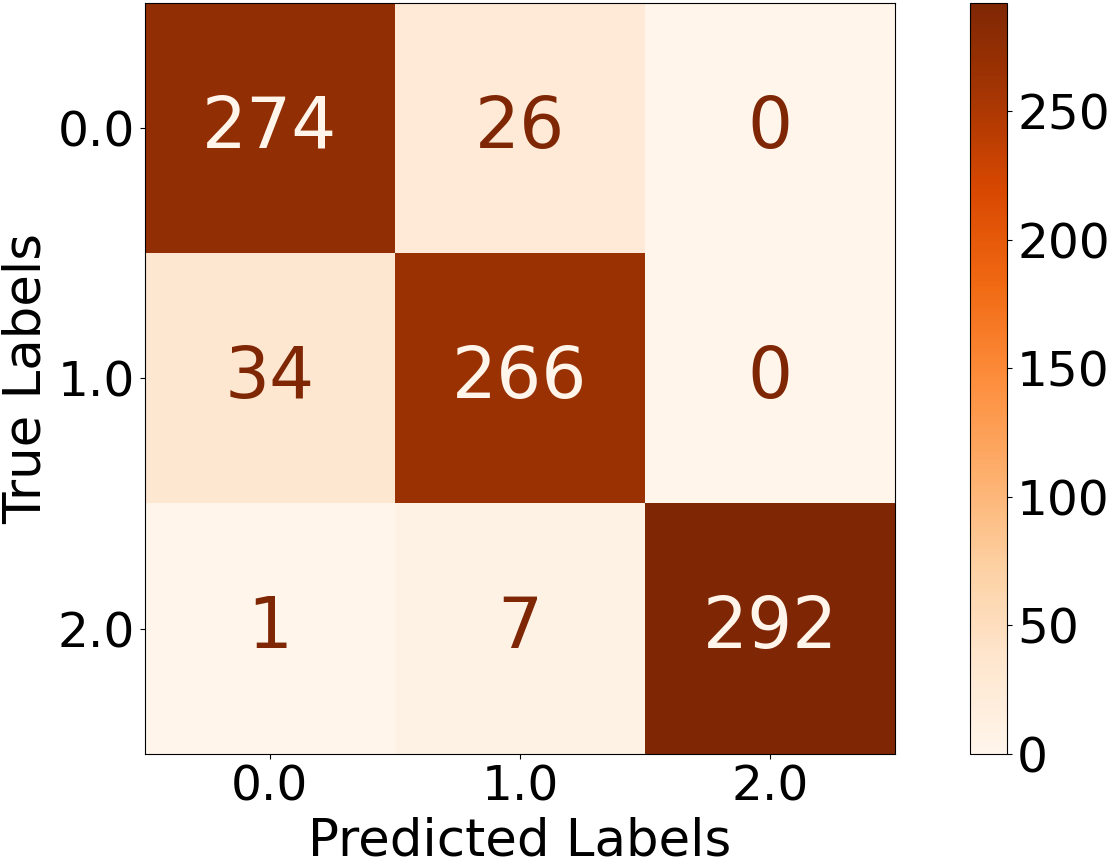}
\caption{}
\label{fig:3_xgb_cm_all_3cls}
\end{subfigure}
\caption{Confusion matrices for 3-class classification of breathing data considering all distances (0.5\,m, 1\,m and 1.5\,m) together found through cross-validation with (a) decision tree, (b) random forest and (c) XGBoost algorithms.}
 \label{fig:3classcm}
\end{figure}

\begin{figure}[!hbt] \centering
\begin{subfigure}[!hbt]{0.24\textwidth}
 \centering
\includegraphics[width=\textwidth]{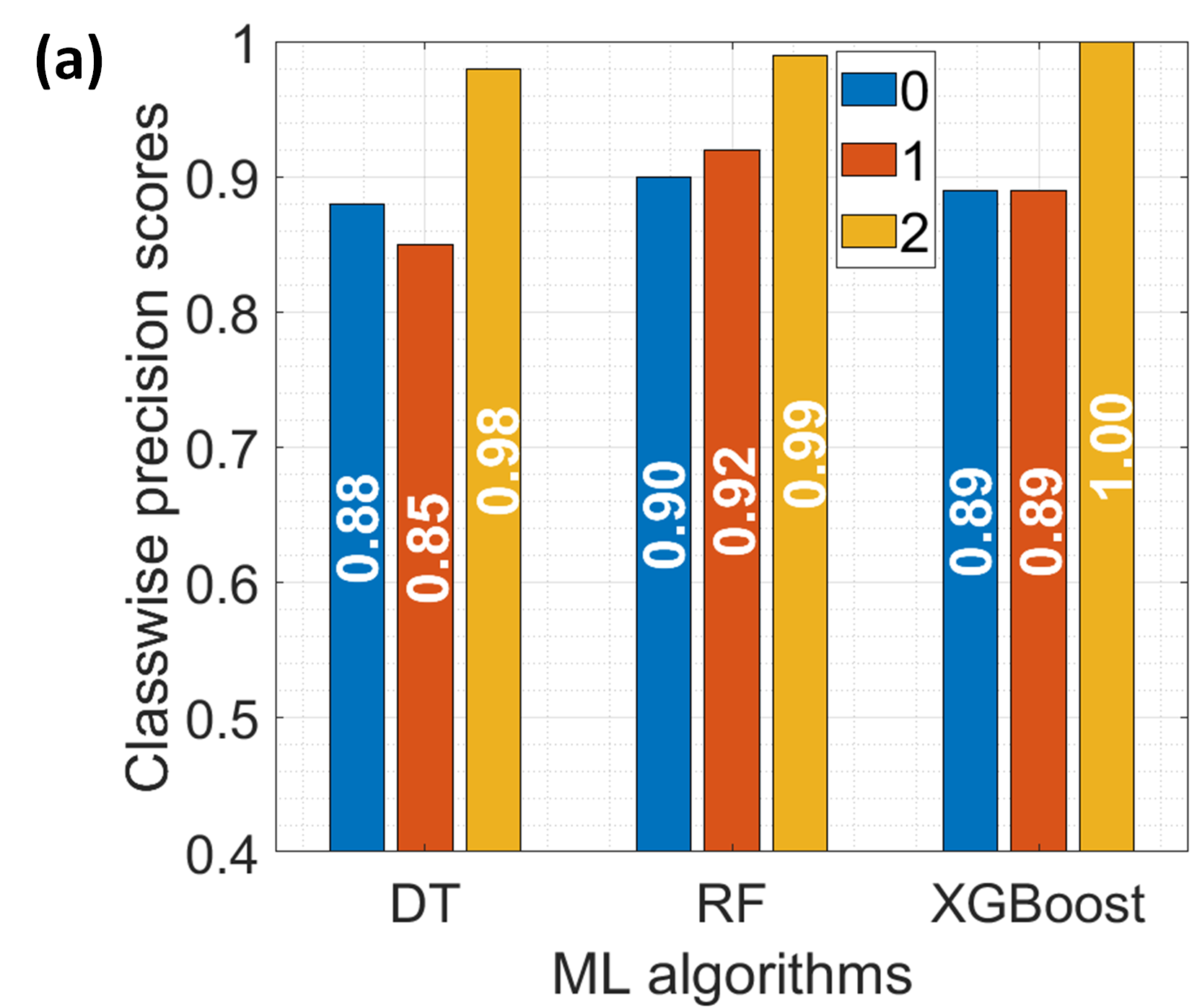}
\label{fig:prec3cls}
\end{subfigure}
\hfill
\begin{subfigure}[!hbt]{0.24\textwidth}
 \centering
\includegraphics[width=\textwidth]{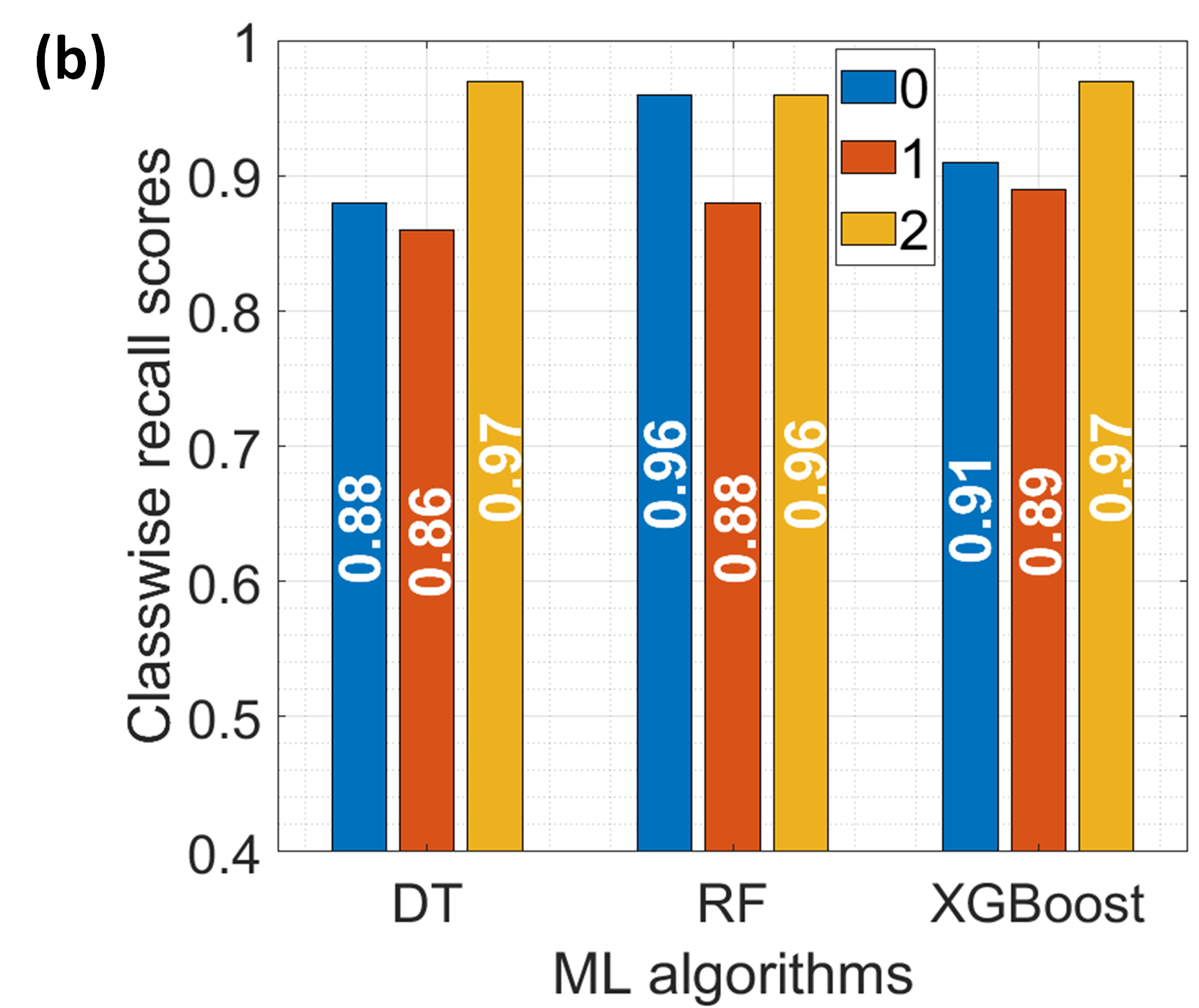}
\label{fig:rec3cls}
\end{subfigure}
\hfill
\caption{Classwise (a) precision and (b) recall plots for 3-class classification with decision tree, random forest and XGBoost models considering all distances (0.5\,m, 1\,m and 1.5\,m) together.}
 \label{fig:3clsbar}
\end{figure}

\begin{figure}[!hbt] \centering
\begin{subfigure}[!hbt]{0.155\textwidth}
 \centering
\includegraphics[width=\textwidth]{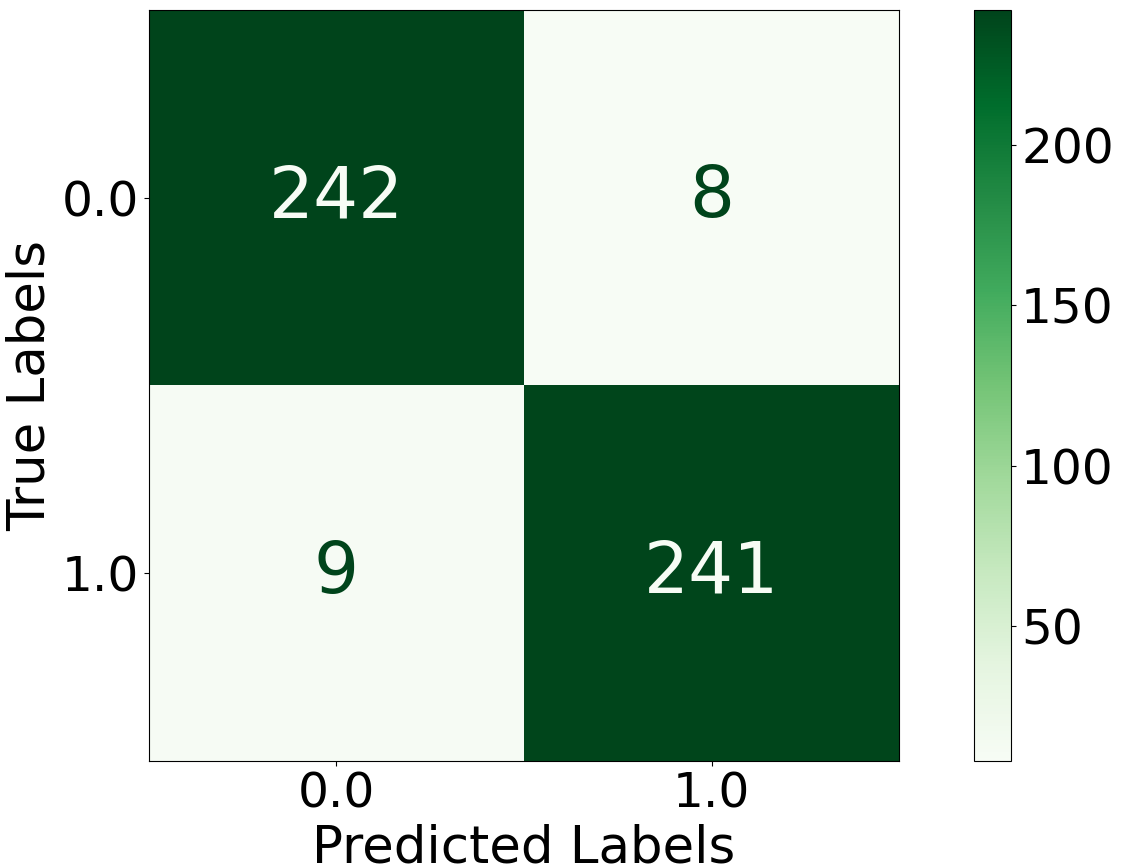}
\caption{}
\label{fig:DT_cm_2cls}
\end{subfigure}
\begin{subfigure}[!hbt]{0.155\textwidth}
 \centering
\includegraphics[width=\textwidth]{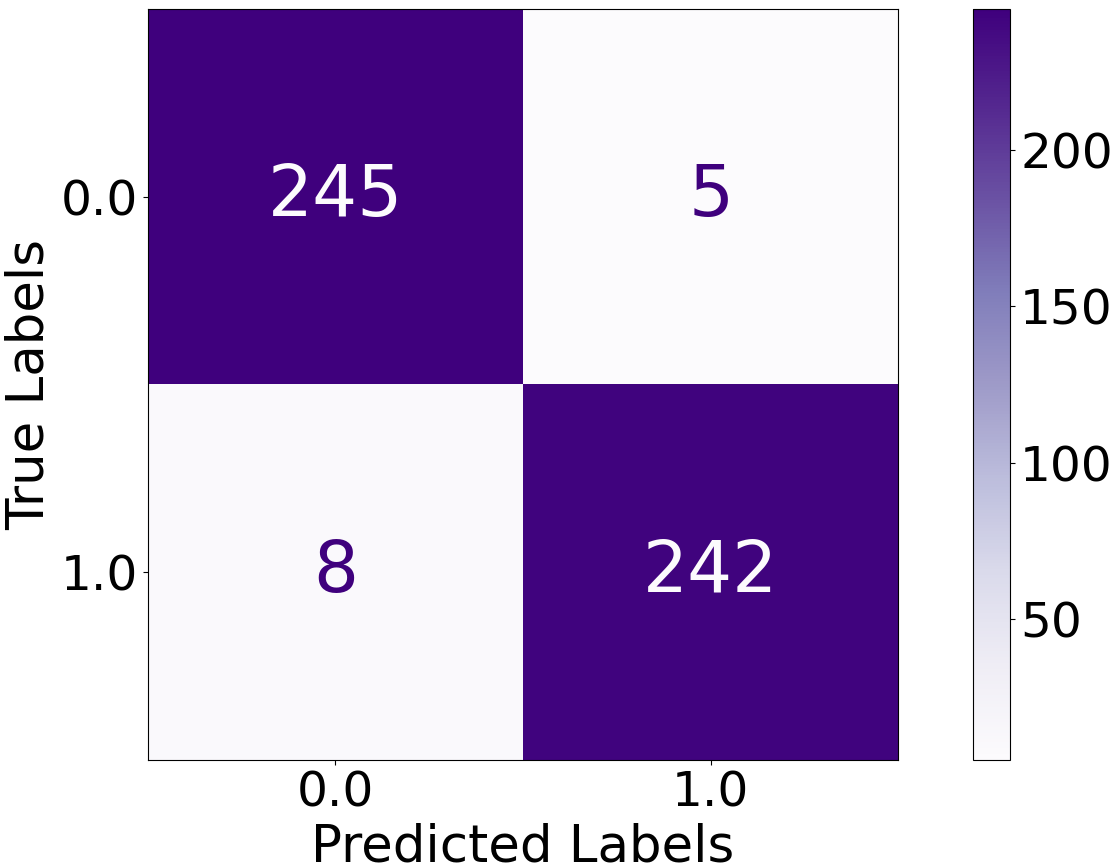}
\caption{}
\label{fig:RF_cm_2cls}
\end{subfigure}
\begin{subfigure}[!hbt]{0.155\textwidth}
 \centering
\includegraphics[width=\textwidth]{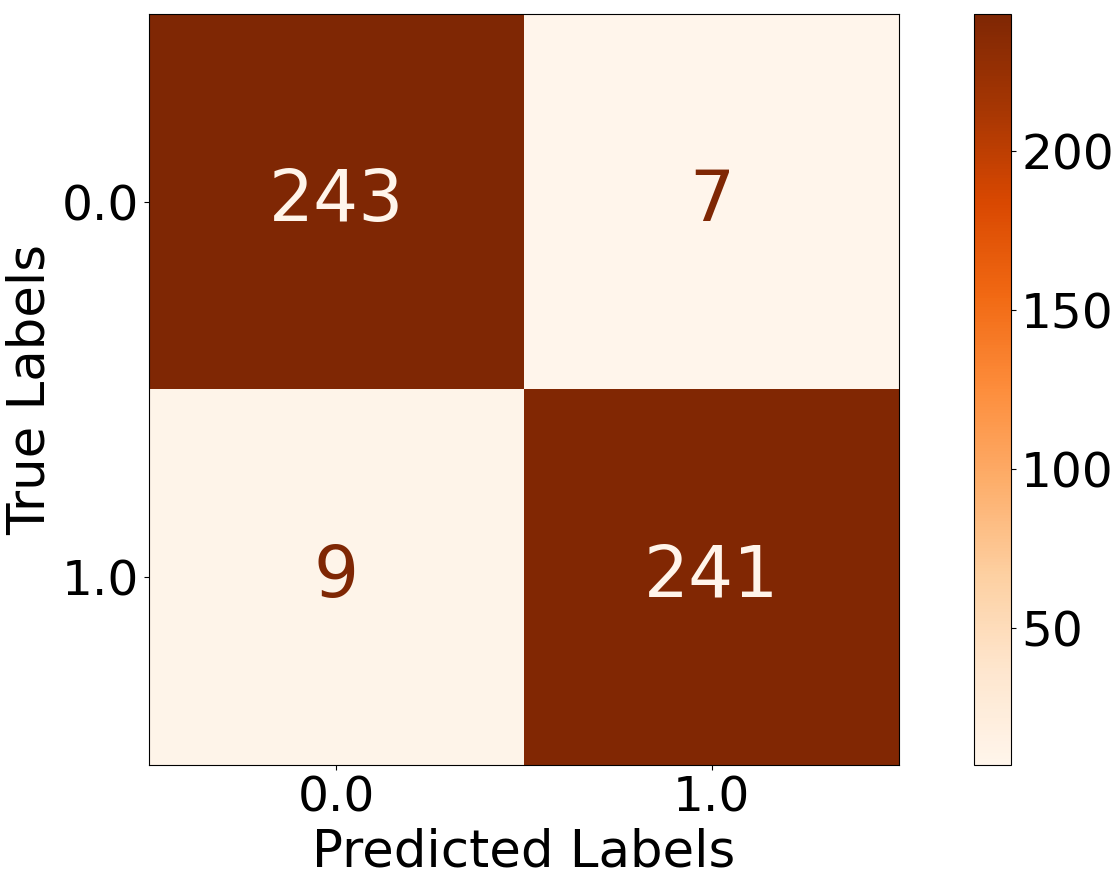}
\caption{}
\label{fig:XGB_cm_2cls}
\end{subfigure}
\caption{Confusion matrices for binary classification of stochastic respiration data considering 0.5 to 2.5\,m distances together found through cross-validation with (a) decision tree, (b) random forest and (c) XGBoost algorithms.}
 \label{fig:stochastic_cm}
\end{figure}

Confusion matrices for 8-class (Fig.~\ref{fig:cm1}), 3-class (Fig.~\ref{fig:3classcm}) and binary classifications (Fig.~\ref{fig:stochastic_cm}) provided a comprehensive view of the classification performance in terms of true positive, true negative, false positive, and false negative counts. 
After analyzing the confusion matrices, several erroneous predictions were observed in cases where breathing rates and depths exhibited values that bordered the range of another class. Specifically, these erroneous predictions were more prevalent in class 5--hyperpnea, which is characterized by lower breathing depth and a normal breathing rate. Instances where the breathing depth closely resembled that of class 0--Eupnea, were inaccurately classified as class 0. Furthermore, instances with a significantly reduced breathing depth, resembling apnea, were occasionally misclassified as class 1--apnea. Similar misclassifications were also observed in class 4 and 6 due to the borderline values of breathing depth with class 0 and 2, respectively. Furthermore, from the diagonally symmetric values in the confusion matrix, it has been understood that analogous misclassifications occurred in the opposite direction too. These observations suggest the challenge posed by borderline values of breathing rates and depths, leading to misclassifications in both directions among adjacent classes.

In both 8-class and 3-class classifications performed, faulty data class was seen to have the highest precision and recall (Fig.~\ref{fig:prec_rec_bar_chart} and~\ref{fig:3clsbar}) which indicates that the system is aptly capable of rejecting the erroneous data which can be commonly generated by body movement, incorrect user posture, and the movement of people nearby. In Fig.~\ref{fig:prec_rec_bar_chart}, the precision and recall values of class 0, 4 and 5 are usually lower than other classes, because of their borderline similarities with each other. From Fig.~\ref{fig:3clsbar}, it is evident that despite of the variance introduced by combining data collected at multiple distances together, both precision and recall values of class 2 (anomalous breathing class) are $\approx 90\%$  with ensemble models like RF and XGBoost. This proves the effectiveness of the system in detecting breathing anomalies with both low false positives and false negatives.

\section{Conclusion and Future Directions}
\label{sec:conclusion}
In conclusion, the infrared light-wave sensing system was able to detect breathing anomalies and erroneous data through machine learning models, with the highest accuracy of 96.75\% obtained by random forest model. Ensemble models were found to be more effective than single decision tree in classifying the datasets collected at various distances. Thus, this study establishes a foundational framework for respiratory anomaly detection using light-wave sensing technology. To enhance the anomaly detection model, incorporating transfer learning by pre-training on a larger dataset representing a broader range of respiratory anomalies and environmental conditions is essential. Additionally, fine-tuning the model to learn the regular breathing patterns specific to individual users will further refine its accuracy. The system can be used by clinicians in hospitals for patient prescreening and continuous monitoring. Its utility extends to non-clinical settings too, offering use cases such as general health monitoring at home, monitoring driver fatigue in vehicles, and assessing the health status of employees in the workplace. 

Key limitations of this study are the use of simulated breathing under highly constrained conditions and the absence of clinical validation. Although stochastic variability of breathing amplitude and rate has been addressed in this study, real human breathing often presents additional irregularities, including variations in inhalation and exhalation durations, as well as occasional pauses. Moreover, the presence of noise and artifacts in the respiration signal originating from physiological (e.g., cardiac motion) and environmental factors could make anomaly detection more challenging in real-world scenarios. Therefore, rigorous clinical trials involving human subjects/patients are imperative to validate the system's robustness and generalizability in realistic environments.


\bibliographystyle{IEEEtran.bst}
\bibliography{references.bib}

\begin{thebibliography}{10}
\providecommand{\url}[1]{#1}
\csname url@samestyle\endcsname
\providecommand{\newblock}{\relax}
\providecommand{\bibinfo}[2]{#2}
\providecommand{\BIBentrySTDinterwordspacing}{\spaceskip=0pt\relax}
\providecommand{\BIBentryALTinterwordstretchfactor}{4}
\providecommand{\BIBentryALTinterwordspacing}{\spaceskip=\fontdimen2\font plus
\BIBentryALTinterwordstretchfactor\fontdimen3\font minus \fontdimen4\font\relax}
\providecommand{\BIBforeignlanguage}[2]{{%
\expandafter\ifx\csname l@#1\endcsname\relax
\typeout{** WARNING: IEEEtran.bst: No hyphenation pattern has been}%
\typeout{** loaded for the language `#1'. Using the pattern for}%
\typeout{** the default language instead.}%
\else
\language=\csname l@#1\endcsname
\fi
#2}}
\providecommand{\BIBdecl}{\relax}
\BIBdecl

\bibitem{elliott2016respiratory}
M.~Elliott, ``{Why is respiratory rate the neglected vital sign? A narrative review},'' \emph{Int Arch Nurs Health Care}, vol.~2, no.~3, p. 050, 2016.

\bibitem{hogan2006don}
J.~Hogan, ``Why don't nurses monitor the respiratory rates of patients?'' \emph{British Journal of Nursing}, vol.~15, no.~9, pp. 489--492, 2006.

\bibitem{van2008missed}
C.~H. Van~Leuvan and I.~Mitchell, ``{Missed opportunities? An observational study of vital sign measurements},'' \emph{Crit Care Resusc}, vol.~10, no.~2, pp. 111--115, 2008.

\bibitem{grover2008automated}
S.~S. Grover and S.~D. Pittman, ``Automated detection of sleep disordered breathing using a nasal pressure monitoring device,'' \emph{Sleep and Breathing}, vol.~12, pp. 339--345, 2008.

\bibitem{Smith2016}
L.~N. Smith, M.~L. Smith, M.~E. Fletcher, and A.~J. Henderson, ``{A 3D machine vision method for non-invasive assessment of respiratory function},'' \emph{International Journal of Medical Robotics and Computer Assisted Surgery}, vol.~12, no.~2, pp. 179--188, Jun 2016.

\bibitem{Abdelnasser2015}
H.~Abdelnasser, K.~A. Harras, and M.~Youssef, ``{UbiBreathe: A ubiquitous non-invasive WiFi-based breathing estimator},'' in \emph{Proceedings of the 16th ACM International Symposium on Mobile Ad Hoc Networking and Computing}, 2015, pp. 277--286.

\bibitem{bao2022wi}
N.~Bao, J.~Du, C.~Wu, D.~Hong, J.~Chen, R.~Nowak, and Z.~Lv, ``Wi-breath: A wifi-based contactless and real-time respiration monitoring scheme for remote healthcare,'' \emph{IEEE journal of biomedical and health informatics}, 2022.

\bibitem{guo2023breatheband}
Z.~Guo, W.~Yuan, L.~Gui, B.~Sheng, and F.~Xiao, ``Breatheband: A fine-grained and robust respiration monitor system using wifi signals,'' \emph{ACM Transactions on Sensor Networks}, vol.~19, no.~4, pp. 1--18, 2023.

\bibitem{saeed2021wireless}
U.~Saeed, S.~Y. Shah, A.~Zahid, J.~Ahmad, M.~A. Imran, Q.~H. Abbasi, and S.~A. Shah, ``{Wireless channel modelling for identifying six types of respiratory patterns with SDR sensing and deep multilayer perceptron},'' \emph{IEEE Sensors Journal}, vol.~21, no.~18, pp. 20\,833--20\,840, 2021.

\bibitem{SanKocur2017}
D.~Kocur, D.~Novák, and J.~Demčák, ``A joint localization and breathing rate estimation of static persons using {UWB} radar,'' in \emph{2017 IEEE International Conference on Systems, Man, and Cybernetics (SMC)}, 2017, pp. 1728--1733.

\bibitem{Purnomo2021}
A.~T. Purnomo, D.-B. Lin, T.~Adiprabowo, and W.~F. Hendria, ``{Non-contact monitoring and classification of breathing pattern for the supervision of people infected by COVID-19},'' \emph{Sensors}, vol.~21, no.~9, p. 3172, 2021.

\bibitem{LAPINSKY2006267}
S.~E. Lapinsky and A.~C. Easty, ``Electromagnetic interference in critical care,'' \emph{Journal of critical care}, vol.~21, no.~3, pp. 267--270, 2006.

\bibitem{MARIAPPAN2016727}
P.~M. Mariappan, D.~R. Raghavan, S.~H.~A. Aleem, and A.~F. Zobaa, ``{Effects of electromagnetic interference on the functional usage of medical equipment by 2G/3G/4G cellular phones: A review},'' \emph{Journal of Advanced Research}, vol.~7, no.~5, pp. 727--738, 2016.

\bibitem{TAKI200140}
M.~Taki and S.~Watanabe, ``Biological and health effects of exposure to electromagnetic field from mobile communications systems,'' \emph{IATSS research}, vol.~25, no.~2, pp. 40--50, 2001.

\bibitem{gye2012effect}
M.~C. Gye and C.~J. Park, ``Effect of electromagnetic field exposure on the reproductive system,'' \emph{Clinical and experimental reproductive medicine}, vol.~39, no.~1, pp. 1--9, 2012.

\bibitem{zhang2023recent}
X.~Zhang, M.~Hu, Y.~Zhang, G.~Zhai, and X.-P. Zhang, ``Recent progress of optical imaging approaches for noncontact physiological signal measurement: A review,'' \emph{Advanced Intelligent Systems}, p. 2200345, 2023.

\bibitem{Brieva2020}
J.~Brieva, H.~Ponce, and E.~Moya-Albor, ``Non-contact breathing rate monitoring system using a magnification technique and convolutional networks,'' in \emph{15th international symposium on medical information processing and analysis}, vol. 11330.\hskip 1em plus 0.5em minus 0.4em\relax SPIE, 2020, pp. 181--189.

\bibitem{romano2021}
C.~Romano, E.~Schena, S.~Silvestri, and C.~Massaroni, ``{Non-contact respiratory monitoring using an RGB camera for real-world applications},'' \emph{Sensors}, vol.~21, no.~15, p. 5126, 2021.

\bibitem{Hwang2021}
H.~Hwang and E.~C. Lee, ``{Non-contact respiration measurement method based on RGB camera using 1D convolutional neural networks},'' \emph{Sensors}, vol.~21, no.~10, 2021.

\bibitem{tan2023lightweight}
X.~Tan, M.~Hu, G.~Zhai, Y.~Zhu, W.~Li, and X.-P. Zhang, ``Lightweight video-based respiration rate detection algorithm: An application case on intensive care,'' \emph{IEEE Transactions on Multimedia}, 2023.

\bibitem{s21041135}
A.~P. Addison, P.~S. Addison, P.~Smit, D.~Jacquel, and U.~R. Borg, ``Noncontact respiratory monitoring using depth sensing cameras: A review of current literature,'' \emph{Sensors}, vol.~21, no.~4, p. 1135, 2021.

\bibitem{kempfle123}
J.~Kempfle and K.~Van~Laerhoven, ``Respiration rate estimation with depth cameras: An evaluation of parameters,'' in \emph{Proceedings of the 5th international Workshop on Sensor-based Activity Recognition and Interaction}, 2018, pp. 1--10.

\bibitem{wang2020unobtrusive}
Y.~Wang, M.~Hu, Y.~Zhou, Q.~Li, N.~Yao, G.~Zhai, X.-P. Zhang, and X.~Yang, ``Unobtrusive and automatic classification of multiple people’s abnormal respiratory patterns in real time using deep neural network and depth camera,'' \emph{IEEE Internet of Things Journal}, vol.~7, no.~9, pp. 8559--8571, 2020.

\bibitem{Jagadev2020}
P.~Jagadev and L.~I. Giri, ``{Non-contact monitoring of human respiration using infrared thermography and machine learning},'' \emph{Infrared Physics and Technology}, vol. 104, 2020.

\bibitem{chen2017hht}
D.-Y. Chen and J.-C. Lai, ``{HHT-based remote respiratory rate estimation in thermal images},'' in \emph{2017 18th IEEE/ACIS International Conference on Software Engineering, Artificial Intelligence, Networking and Parallel/Distributed Computing (SNPD)}.\hskip 1em plus 0.5em minus 0.4em\relax IEEE, 2017, pp. 263--268.

\bibitem{shu2022}
S.~Shu, H.~Liang, Y.~Zhang, Y.~Zhang, and Z.~Yang, ``Non-contact measurement of human respiration using an infrared thermal camera and the deep learning method,'' \emph{Measurement Science and Technology}, vol.~33, no.~7, p. 075202, 2022.

\bibitem{Abuella2020}
H.~Abuella and S.~Ekin, ``Non-contact vital signs monitoring through visible light sensing,'' \emph{IEEE Sensors Journal}, vol.~20, no.~7, pp. 3859--3870, 2020.

\bibitem{Rehman2021}
M.~Rehman, R.~A. Shah, M.~B. Khan, S.~A. Shah, N.~A. Abuali, X.~Yang, A.~Alomainy, M.~A. Imran, and Q.~H. Abbasi, ``{Improving machine learning classification accuracy for breathing abnormalities by enhancing dataset},'' \emph{Sensors}, vol.~21, no.~20, pp. 1--15, 2021.

\bibitem{Ucar2017}
M.~K. U{\c{c}}ar, M.~R. Bozkurt, C.~Bilgin, and K.~Polat, ``{Automatic detection of respiratory arrests in OSA patients using PPG and machine learning techniques},'' \emph{Neural Computing and Applications}, vol.~28, no.~10, pp. 2931--2945, 2017.

\bibitem{Pegoraro2021}
J.~A. Pegoraro, S.~Lavault, N.~Wattiez, T.~Similowski, J.~Gonzalez-Bermejo, and E.~Birmel{\'{e}}, ``{Machine-learning based feature selection for a non-invasive breathing change detection},'' \emph{BioData Mining}, vol.~14, no.~1, Dec 2021.

\bibitem{Fekr2016}
A.~R. Fekr, M.~Janidarmian, K.~Radecka, and Z.~Zilic, ``{Respiration Disorders Classification with Informative Features for m-Health Applications},'' \emph{IEEE Journal of Biomedical and Health Informatics}, vol.~20, no.~3, pp. 733--747, May 2016.

\bibitem{Kim2019}
S.~H. Kim, Z.~W. Geem, and G.~T. Han, ``{A novel human respiration pattern recognition using signals of ultra-wideband radar sensor},'' \emph{Sensors (Switzerland)}, vol.~19, no.~15, 2019.

\bibitem{BarbosaPereira2017}
C.~{Barbosa Pereira}, X.~Yu, M.~Czaplik, V.~Blazek, B.~Venema, and S.~Leonhardt, ``{Estimation of breathing rate in thermal imaging videos: a pilot study on healthy human subjects},'' \emph{Journal of Clinical Monitoring and Computing}, vol.~31, no.~6, pp. 1241--1254, 2017.

\bibitem{Yuan2013}
G.~Yuan, N.~A. Drost, and R.~A. McIvor, ``Respiratory rate and breathing pattern,'' \emph{McMaster Univ. Med. J}, vol.~10, no.~1, pp. 23--25, 2013.

\bibitem{Parreira2010}
V.~F. Parreira, C.~J. Bueno, D.~C. Fran{\c{c}}a, D.~S. Vieira, D.~R. Pereira, and R.~R. Britto, ``Breathing pattern and thoracoabdominal motion in healthy individuals: influence of age and sex,'' \emph{Brazilian Journal of Physical Therapy}, vol.~14, pp. 411--416, 2010.

\bibitem{Ali2021}
M.~Ali, A.~Elsayed, A.~Mendez, Y.~Savaria, and M.~Sawan, ``Contact and remote breathing rate monitoring techniques: A review,'' \emph{IEEE Sensors Journal}, vol.~21, no.~13, pp. 14\,569--14\,586, 2021.

\bibitem{Moraes2019}
A.~G. de~Moraes and S.~Surani, ``{Effects of diabetic ketoacidosis in the respiratory system},'' \emph{World Journal of Diabetes}, vol.~10, no.~1, pp. 16--22, 2019.

\bibitem{Leung2012}
R.~S. Leung, V.~R. Comondore, C.~M. Ryan, and D.~Stevens, ``{Mechanisms of sleep-disordered breathing: Causes and consequences},'' pp. 213--230, 2012.

\bibitem{Weinreich2009}
G.~Weinreich, J.~Armitstead, V.~T{\"{o}}pfer, Y.~M. Wang, Y.~Wang, and H.~Teschler, ``{Validation of ApneaLink as screening device for Cheyne-Stokes respiration},'' \emph{Sleep}, vol.~32, no.~4, pp. 553--557, 2009.

\bibitem{gfeller1979}
F.~Gfeller and U.~Bapst, ``Wireless in-house data communication via diffuse infrared radiation,'' \emph{Proceedings of the IEEE}, vol.~67, no.~11, pp. 1474--1486, 1979.

\bibitem{pathak2015}
P.~H. Pathak, X.~Feng, P.~Hu, and P.~Mohapatra, ``Visible light communication, networking, and sensing: A survey, potential and challenges,'' \emph{IEEE Communications Surveys \& Tutorials}, vol.~17, no.~4, pp. 2047--2077, 2015.

\bibitem{detector2}
\BIBentryALTinterwordspacing
{Si Free-space Amplified Photodetectors}. Thorlabs, Inc. Accessed: Apr. 3, 2024. [Online]. Available: \url{https://www.thorlabs.com/catalogpages/Obsolete/2018/PDA100A.pdf}
\BIBentrySTDinterwordspacing

\bibitem{lockin1}
\BIBentryALTinterwordspacing
{Model SR830 DSP Lock-In Amplifier}. Stanford Research Systems. Accessed: Apr. 3, 2024. [Online]. Available: \url{https://www.thinksrs.com/downloads/pdfs/manuals/SR830m.pdf}
\BIBentrySTDinterwordspacing

\bibitem{Breathing_Model}
K.~Y. Gavrilov and T.~Y. Shevgunov, ``A new model of human respiration for algorithm simulation modeling in radar applications,'' in \emph{2020 Systems of Signals Generating and Processing in the Field of on Board Communications}, 2020, pp. 1--5.

\bibitem{terabee}
\BIBentryALTinterwordspacing
{TeraRanger Evo 60m}. Terabee. Accessed: Apr. 3, 2024. [Online]. Available: \url{https://www.mouser.com/datasheet/2/944/TeraRanger-Evo-60m-Specification-sheet-3-1729032.pdf}
\BIBentrySTDinterwordspacing

\bibitem{transmitter}
\BIBentryALTinterwordspacing
{Total Invisible 940nm IR lamp Board with Light Sensor (48 Black LED Illuminator Array) 30ft Range, 120 deg, 12VDC}. Amazon.com, Inc. Accessed: Apr. 3, 2024. [Online]. Available: \url{https://www.amazon.com/gp/product/B0785W2RQQ}
\BIBentrySTDinterwordspacing

\bibitem{funcgen}
\BIBentryALTinterwordspacing
\emph{33500B and 33600A Series Trueform Waveform Generators (20, 30, 80, 120MHz)}, Keysight Technologies, accessed: Apr. 3, 2024. [Online]. Available: \url{https://www.keysight.com/us/en/assets/7018-05928/data-sheets/5992-2572.pdf}
\BIBentrySTDinterwordspacing

\bibitem{adc}
\BIBentryALTinterwordspacing
{DAQC2plate - Affordable high precision multifunctional Data Acquisition and Controller board.} Pi-Plates. Accessed: Apr. 3, 2024. [Online]. Available: \url{https://pi-plates.com/daqc2r1/}
\BIBentrySTDinterwordspacing

\bibitem{acute}
T.~R. Gravelyn and M.~John G.~Weg, ``Respiratory rate as an indicator of acute respiratory dysfunction,'' \emph{The Journal of the American Medical Association}, vol. 244, no.~10, pp. 1123--1125, 1980.

\bibitem{Lee2020}
\BIBentryALTinterwordspacing
T.-H. Lee, A.~Ullah, and R.~Wang, \emph{Bootstrap Aggregating and Random Forest}.\hskip 1em plus 0.5em minus 0.4em\relax Cham: Springer International Publishing, 2020, pp. 389--429. [Online]. Available: \url{https://doi.org/10.1007/978-3-030-31150-6_13}
\BIBentrySTDinterwordspacing

\bibitem{sagi2018ensemble}
O.~Sagi and L.~Rokach, ``Ensemble learning: A survey,'' \emph{Wiley Interdisciplinary Reviews: Data Mining and Knowledge Discovery}, vol.~8, no.~4, p. e1249, 2018.

\end{thebibliography}

\end{document}